\documentclass[%
 preprint,
 amsmath,amssymb,
 aps,
prl,
]{revtex4-2}

\usepackage{graphicx}
\usepackage{dcolumn}
\usepackage{bm}
\usepackage[hidelinks]{hyperref}
\usepackage{color} 
\usepackage{siunitx}
\DeclareSIUnit\eangstrom{\text{e\AA}}
\usepackage{amsmath}
\usepackage{mathtools}
\usepackage[version=4]{mhchem} 
\DeclarePairedDelimiter\abs{\lvert}{\rvert}
\newcommand{\chiint}{\chi_\text{int}}
\newcommand{\chitrs}{\chi_\text{TRS}}
\newcommand{\eqt}{\hspace{0.17em}{=}\hspace{0.17em}}
\newcommand{\bk}{\mathbf{k}}
 
\newcommand{\delE}{\mathbf{\Delta}E_{\tau,\alpha}}
\newcommand{\deltagap}{\Delta_\text{gap}^{\pm \text{K}}}
\newcommand{\detuning}{\Delta_\omega}
\usepackage[normalem]{ulem}
\usepackage{fixltx2e} 
\begin{document}


\section{Title}
Direct measurement of broken time-reversal symmetry in centrosymmetric and non-centrosymmetric atomically thin crystals with nonlinear Kerr rotation

\section{Author list}
Florentine Friedrich$^{1,\dagger}$, Paul Herrmann$^{1,\dagger}$, Shridhar Sanjay Shanbhag$^2$, Sebastian Klimmer$^{1,3}$, Jan Wilhelm$^{2}$, and {Giancarlo} Soavi$^{1,4,\star}$

\section{Affiliations}
\noindent
$^1$Institute of Solid State Physics, Friedrich Schiller University Jena, Helmholtzweg 5, 07743 Jena, Germany
\newline
$^2$Institute of Theoretical Physics and Regensburg Center for Ultrafast Nanoscopy (RUN), University of Regensburg, Universitätsstrasse 31, 93053 Regensburg, Germany
\newline
$^3$ARC Centre of Excellence for Transformative Meta-Optical Systems, Department of Electronic Materials Engineering, Research School of Physics,
The Australian National University, Canberra ACT 2601, Australia
\newline
$^4$Abbe Center of Photonics, Friedrich Schiller University Jena, Albert-Einstein-Straße 6, 07745 Jena, Germany
\newline
$^{\dagger}$ Both authours contributed equally.
$^{\star}$ giancarlo.soavi@uni-jena.de

\maketitle

\section{Abstract}

Time-reversal symmetry, together with space-inversion symmetry, is one of the defining properties of crystals, underlying phenomena such as magnetism, topology and non-trivial spin textures. Transition metal dichalcogenides (TMDs) provide an excellent tunable model system to study the interplay between time-reversal and space-inversion symmetry, since both can be engineered on demand by tuning the number of layers and \textit{via} all-optical bandgap modulation. In this work, we modulate and study time-reversal symmetry using third harmonic Kerr rotation in mono- and bilayer TMDs. By illuminating the samples with elliptically polarized light, we achieve spin-selective bandgap modulation and consequent breaking of time-reversal symmetry. The reduced symmetry modifies the nonlinear susceptibility tensor, causing a rotation of the emitted third harmonic polarization. With this method, we are able to probe broken time-reversal symmetry in both non-centrosymmetric (monolayer) and centrosymmetric (bilayer) crystals. Furthermore, we discuss how the detected third harmonic rotation angle directly links to the spin-valley locking in monolayer TMDs and to the spin-valley-layer locking in bilayer TMDs. Thus, our results define a powerful approach to study broken time-reversal symmetry in crystals regardless of space-inversion symmetry, and shed light on the spin, valley and layer coupling of atomically thin semiconductors.

\section{Introduction}
Time-reversal symmetry (TRS), in combination with space-inversion symmetry (SIS), defines the energy-spin properties and the Berry curvature of crystals~\cite{soavi2025signaturetopologypolarchiral}. Thus, the presence or absence of TRS and SIS is paramount to explain and understand a variety of phenomena in condensed matter physics, ranging from magnetism~\cite{Fiebig2023} to spin-valley locking~\cite{PhysRevLett.108.196802,Mak2012}, anomalous transport~\cite{KLITZING1983525,McIver2020}, and topology~\cite{Bao2022}. Nonlinear optics (NLO) is an established and powerful tool to investigate the effect of TRS breaking in crystals~\cite{Fiebig2023}. The most relevant example is arguably the study of ferroic materials~\cite{Fiebig2023}, where a vast variety of physical and optical properties is captured by NLO effects described by the magnetic point groups~\cite{Guccione1963}, also known by the name of black and white, Shubnikov, Heesch or Opechowski-Guccione groups~\cite{Lifshitz2005}. Among the possible applications of NLO to the study of time-noninvariant phenomena, second harmonic (SH) nonlinear Kerr rotation certainly plays a key role. It has been predicted and experimentally observed that nonlinear Kerr rotation offers higher sensitivity compared to its linear counterpart~\cite{PUSTOGOWA1995269,PhysRevLett.74.3692}, and the technique has already been applied to bulk~\cite{ZVEZDIN1997444,PhysRevLett.92.047401} and atomically thin crystals~\cite{Wu2023,10.1063/1.372586}. However, most measurements to date were limited to non-centrosymmetric crystals, while the measurement of broken TRS in centrosymmetric crystals remains a challenge. For this scenario, the most common approach to study broken TRS relies on the second order electric quadrupole response~\cite{Ahn2024}, which is orders of magnitude weaker compared to the electric dipole NLO signals and it requires multiple polarization scans at oblique angles of incidence for a full characterization. Using engineered topological optical fields, it has been shown recently that room-temperature realization of valleytronics in centrosymmetric systems is possible~\cite{Mrudul2021, Mitra2024, Tyulnev2024}.

\begin{figure}[]
    \centering
    \includegraphics[width=\linewidth]{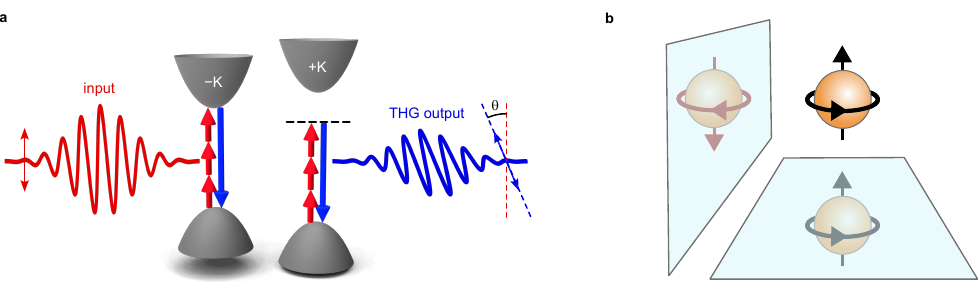}
    \caption{\textbf{Principle of TH Kerr rotation and mirror symmetries of a rotating sphere.} \textbf{a} For a monolayer TMD, broken TRS is equivalent to an asymmetry between the $\pm$K valleys. TRS breaking introduces new $\boldsymbol{\chi^{(3)}}$ tensor elements, leading to a rotation of the TH output by the angle $\theta$. \textbf{b} A rotating sphere as analogue for spin-oriented states. A sphere rotating counter-clockwise (indicated by the black circular arrow and straight black arrow pointing upwards) coincides with its mirror image when mirroring along a horizontal plane (see bottom), but the rotation direction changes for the mirror image when mirroring along a vertical plane (see left, indicated by the red circular arrow and straight red arrow pointing downwards). To keep the rotating sphere invariant under vertical mirroring, the rotation direction needs to be reversed (antisymmetry operation). Figure adapted from Ref.~\cite{Fecher2022}.}
    \label{fig:FIG1_NL_Kerr_rotating_sphere}
\end{figure}

Here, we propose a different approach to study broken TRS in any system, both centrosymmetric and non-centrosymmetric, based on third harmonic (TH) Kerr rotation. To clarify the operating principle of TH Kerr rotation, we refer to the sketch in  Fig.~\ref{fig:FIG1_NL_Kerr_rotating_sphere}a, which depicts the case of a monolayer TMD with broken SIS. When TRS is preserved, the TH signal is parallel to the input fundamental beam (FB). For broken TRS, \textit{i.e.} when there is an asymmetry between the $\pm$K points, the TH signal is rotated with respect to the FB, and the degree of broken TRS is proportional to the rotation angle $\theta$. With this simple and direct approach, we detect broken TRS in both monolayer and bilayer TMDs, namely in both non-centrosymmetric and centrosymmetric crystals. We interpret the results both classically, based on the differences between the NLO susceptibilities of time-invariant (crystallographic) and magnetic point groups, and from an analytical model based on the semiconductor Bloch equations (SBE) for the $\boldsymbol{\chi^{(3)}}$ tensor. With both approaches, we show that broken TRS is induced by off-resonant excitation with circularly polarized light which is responsible for a spin-selective bandgap opening. This spin-selective off-resonant excitation has different effects in monolayer compared to bilayer TMD samples. In fact, in monolayer TMDs the broken SIS leads to spin-valley locking. Thus, off-resonant circularly polarized light creates a valley imbalance by the valley-exclusive optical Stark and Bloch-Siegert effects~\cite{Sie2017,Kim2014}. In contrast, in bilayer TMDs, where SIS is preserved, the $\pm$K valleys are spin-degenerate, but the same spin couples to opposite layers in the opposite valleys (spin-valley-layer locking)~\cite{Jones2014}. In this case, spin-selective excitation by circularly polarized light does not create a valley imbalance, but it simultaneously lifts the spin degeneracy in both valleys. This corresponds to an engineered band dispersion where TRS is broken while SIS is preserved. Thus, our results not only define a powerful approach to study broken TRS in both centrosymmetric and non-centrosymmetric crystals, but they also provide a unique tool to investigate the fascinating spin, valley and layer coupling of atomically thin semiconductors, which represent a platform for ultrafast valleytronic logic operations~\cite{Gucci2024}. 

\section{Handedness and magnetic point groups}

To understand the symmetry operations that define the time-invariant and time-noninvariant terms of the NLO susceptibilities in mono- and bilayer TMDs, we first need to discuss the property of \textit{handedness} of spin-oriented states. As an example, we take a rotating sphere, as depicted in Fig.~\ref{fig:FIG1_NL_Kerr_rotating_sphere}b. The sphere possesses the binary property of \textit{handedness}: it rotates either right (clockwise or $\uparrow$) or left (counter-clockwise or $\downarrow$). When the rotating sphere is mirrored along a horizontal plane (perpendicular to the axis of rotation), the rotation remains unchanged. Thus, this horizontal reflection is a symmetry operation. For mirror operations along a vertical plane (parallel to the axis of rotation), the rotation direction changes from right to left. Therefore, also the arrow is flipped from up to down~\cite{Fecher2022}. This vertical reflection can be considered as a symmetry operation only in combination with a flip of the rotation direction, \textit{i.e.} an antisymmetry operation~\cite{Padmanabhan2020}. In general, symmetry operations within a unit cell of a crystal can be grouped into five categories: identity, proper rotation, reflection, inversion and roto-reflection. By combining these symmetry operations, one can derive the \textbf{crystallographic} point groups, which ultimately define the NLO properties of crystals as long as TRS is preserved~\cite{Fiebig2023}. Following the discussion above, we can now consider a binary information on each site of the unit cell, such as the electron spin. This binary property can be reversed by the antisymmetry operation, \textit{i.e.} by changing spin up to spin down (or analogously by changing \textit{black} into \textit{white}~\cite{Padmanabhan2020}). Combining \textbf{crystallographic} point groups with this binary property defines the \textbf{magnetic} (\textit{black} and \textit{white}) point groups. These are fundamental to describe crystal properties upon breaking of TRS~\cite{Lifshitz2005}. In the following we will summarize symmetry operations in mono- and bilayer TMDs where TRS is either preserved or broken. For this, we will use the stereographic projections of their magnetic point group, which provide an intuitive way to graphically represent all allowed symmetry operations (see Supplementary Information section S4 for further details).

\section{Symmetry of mono- and bilayer TMDs}

TMDs are the ideal platform to investigate the interplay between SIS and TRS. Both properties can be engineered independently by tuning the number of layers (in the 2H phase, an even/odd number of layers possesses/breaks SIS)~\cite{Zhang2014} and \textit{via} interaction with light~\cite{Sie2017,Kim2014}. In addition, layered materials display a strong NLO response~\cite{https://doi.org/10.1002/lpor.202100726} despite their atomic thickness, which enables a variety of applications ranging from all-optical~\cite{Klimmer2021} and electrical tunability~\cite{https://doi.org/10.1002/advs.202401840,Soavi2018}, to ultrafast logic operations~\cite{Li2022,Gucci2024}, gas sensing~\cite{An2020} and nonlinear valleytronics~\cite{https://doi.org/10.1002/smll.202301126}. Thus, layered materials are ideal systems to explore the potentials of TH Kerr rotation. 

\begin{figure}
    \centering
    \includegraphics[width=\linewidth]{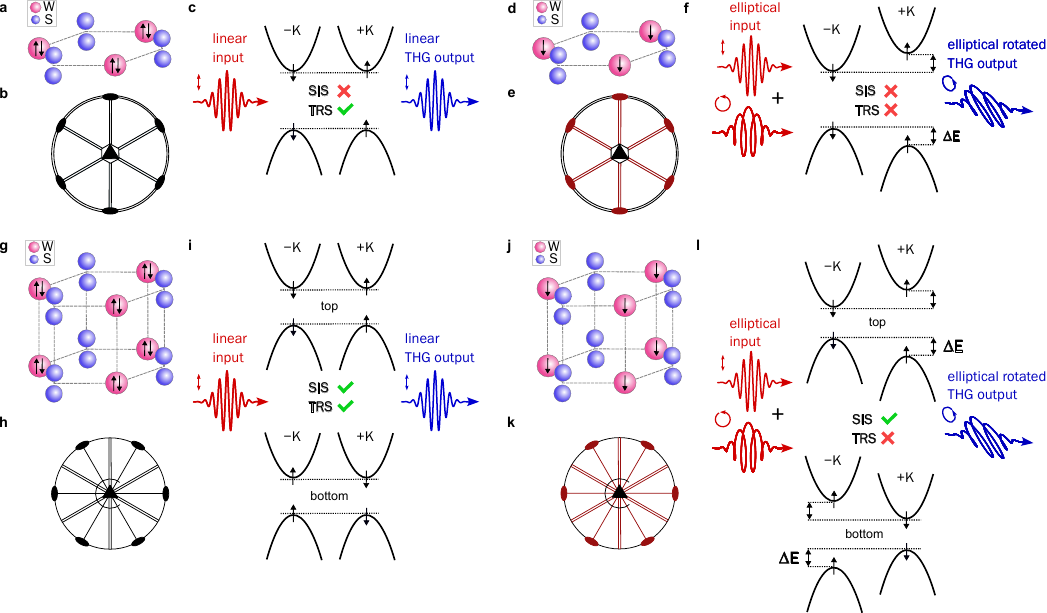}
    \caption{\textbf{Probing TRS breaking in mono- and bilayer \ce{WS_2} \textit{via} TH Kerr rotation.} \textbf{a} Real-space schematic of a \ce{WS_2} monolayer with preserved SIS and TRS. \textbf{b} Stereographic projection of the $D_{3h} \equiv \bar6m2$ crystallographic point group. \textbf{c} Energy and spin of the $\pm$K valleys of a \ce{WS_2} monolayer with preserved TRS. Linear excitation leads to linear TH parallel to the input. \textbf{d} Real-space schematic of a \ce{WS_2} monolayer with broken SIS and TRS. Due to the spin-selective gap opening, one spin dominates. \textbf{e} Stereographic projection of the $\bar6m'2'$ magnetic point group. Red elements represent symmetry operations that are only allowed under antisymmetry. \textbf{f} Energy and spin of the $\pm$K valleys of a \ce{WS_2} monolayer with broken TRS. Elliptical excitation leads to the opening of a gap at the $+$K valley by $\mathbf{\Delta} E$ and a rotated TH output. \textbf{g} Real-space schematic of a \ce{WS_2} bilayer with preserved SIS and TRS. \textbf{h} Stereographic projection of the $D_{3d} \equiv \bar3m$ crystallographic point group. \textbf{i} Energy and spin of the $\pm$K valleys of a \ce{WS_2} bilayer with preserved TRS. Within one layer, the spin states differ for each valley because of spin-valley-layer locking. Linear excitation leads to linear TH parallel to the input. \textbf{j} Real-space schematic of a \ce{WS_2} bilayer with preserved SIS, but broken TRS. Due to spin-selective gap opening one of the spins is now dominant. \textbf{k} Stereographic projection of the $\bar3m'$ magnetic point group. \textbf{l} Energy and spin of the $\pm$K valleys of a \ce{WS_2} bilayer with broken TRS. Elliptical excitation leads to the opening of a gap by $\mathbf{\Delta} E$ in the top and bottom layers and thus to a rotated TH output.} 
    \label{fig:FIG2_schematics_mono_bi}
\end{figure}

Monolayer TMDs are non-centrosymmetric crystals and belong to the $D_{3h}$ crystallographic point group in the Schoenflies notation, or $\bar6m2$ in the Hermann-Mauguin notation. They are direct gap semiconductors with optical transitions at the $\pm$K valleys of the Brillouin zone. Since the $\pm$K transitions are mainly defined by the \textit{d}-orbitals of the transition metal atoms~\cite{Liu2015}, we represent the spin states of the lowest energy transitions on the W sites in Fig.~\ref{fig:FIG2_schematics_mono_bi}a. Broken SIS combined with preserved TRS imposes that $E_\uparrow (+\bk) \neq E_\uparrow (-\bk)$ and $E_\uparrow (+\bk) = E_\downarrow (-\bk)$, leading to the so-called spin-valley locking~\cite{Xu2014}. Breaking of TRS in monolayer TMDs is equivalent to lifting the energy degeneracy between the $\pm$K valleys. The stereographic projection in Fig.~\ref{fig:FIG2_schematics_mono_bi}b shows a summary of the allowed symmetry operations in a monolayer TMD when TRS is preserved (see Supplementary Information section S4 for details). When TRS is broken, the symmetry of monolayer TMDs reduces to the $\bar6m'2'$ magnetic point group (see Fig.~\ref{fig:FIG2_schematics_mono_bi}d): the horizontal mirror symmetry and three-fold rotation symmetry remain allowed without having to perform the antisymmetry operation (thus still depicted in black in Fig.~\ref{fig:FIG2_schematics_mono_bi}e), while the $C_2$ and vertical mirror symmetry are only allowed in combination with the antisymmetry operation (thus depicted in red). To understand this, we refer again to the example of the rotating sphere (equivalent to the spin) in Fig.~\ref{fig:FIG1_NL_Kerr_rotating_sphere}b.

In contrast, AB-stacked bilayer TMDs are centrosymmetric crystals with $D_{3d}$ ($\bar3m$) symmetry when TRS is preserved. Bilayer TMDs are indirect gap semiconductors~\cite{Splendiani2010}, but they still have a direct-gap transition at the $\pm $K valleys. Since both SIS and TRS are preserved, the $\pm$K transition must be spin-degenerate. In each valley, opposite spins are coupled to opposite layers (see Fig.~\ref{fig:FIG2_schematics_mono_bi}i), leading to the so-called spin-valley-layer locking~\cite{Jones2014}. Breaking of TRS reduces the symmetry of bilayer TMDs to the $\bar3m'$ magnetic point group: vertical mirror planes and the $C_2$ symmetry operations are only allowed in combination with the antisymmetry operation (see the stereographic projection in Fig.~\ref{fig:FIG2_schematics_mono_bi}k and Supplementary Information section S4 for further details). We stress that breaking of TRS in bilayer TMDs is not equivalent to lifting the energy degeneracy between the valleys, but rather to a split between the spin up/down energy levels in both valleys and in opposite layers (see Fig.~\ref{fig:FIG2_schematics_mono_bi}l). This effect is responsible for a large degree of circular polarization in photoluminescence (PL) experiments, which was observed even at room temperature in \ce{WS_2} and \ce{WSe_2}~\cite{nayak2016robust,doi:10.1021/acs.jpclett.1c01578}.

Based on these symmetry considerations, we will now demonstrate that broken TRS can be easily detected as a rotation in the TH polarization with respect to the polarization of the FB in both mono- and bilayer TMDs. 

\section{Analytical model and experimental results}{\label{sec:experimental_results}}
\subsection{TH Kerr rotation in monolayer WS\textsubscript{2}}

First, we study the non-centrosymmetric \ce{WS_2} monolayer. The excitation of TMD monolayers with circularly polarized light is both spin and valley selective~\cite{Herrmann2025}. In contrast, linearly polarized light interacts equally with both valleys. Thus, when the FB is linearly polarized and TRS is preserved, the emitted TH signal is linearly polarized and parallel to the input FB, as depicted in Fig.~\ref{fig:FIG2_schematics_mono_bi}c. The situation changes drastically when an elliptically polarized pulse interacts with the sample. To understand this, we start from a simple classical description of third harmonic generation (THG). The electric field generated by elliptically polarized light of frequency $\omega$ with a field strength $\mathcal{E}$ and ellipticity angle $\alpha$ is given by: 

\begin{align}
\boldsymbol{\mathcal{E}}(t)=\mathcal{E}
\left( 
\begin{array}{c}
\cos\alpha\cos \omega t \\
\sin\alpha\sin \omega t
\end{array}
\right).\label{e1}
\end{align}

The elliptical polarization can be interpreted as a superposition of linearly and circularly polarized light. While the linearly polarized part has no effect on TRS, the circularly polarized component of the FB creates a valley imbalance by breaking TRS \textit{via} the valley-selective optical Stark (OS) and Bloch-Siegert (BS) effects~\cite{Herrmann2025,Sie2017}, see Fig.~\ref{fig:FIG2_schematics_mono_bi}f. We represent this, in real space, by the presence of a spin-oriented state (corresponding to the lowest energy transition in the $-$K valley) on the W atoms in Fig.~\ref{fig:FIG2_schematics_mono_bi}d. In this new configuration, the symmetry of the monolayer is reduced to $\bar6m'2'$, as discussed above and depicted in Fig.~\ref{fig:FIG2_schematics_mono_bi}e. 

If we assume normal incidence of the electromagnetic field on the sample, then the $\bar6m2$ crystallographic point group has the following nonzero in-plane elements of the $\boldsymbol{\chi^{(3)}}$ tensor~\cite{Boyd2020}:

\begin{align*}
    \chi^{(3)}_{xxxx} &= \chi^{(3)}_{yyyy} = 3\chi^{(3)}_{xxyy} =3 \chi^{(3)}_{xyyx} =3 \chi^{(3)}_{xyxy} \equiv \chiint
\end{align*}

where we refer to $ \chiint$ as the intrinsic (time-invariant) element of the NLO susceptibility. It is trivial to check that, in this case, the polarization of the TH signal is always parallel to that of the input FB. Upon breaking of TRS, in addition to these terms, the $\bar6m'2'$ magnetic point group has the following non-zero elements of the $\boldsymbol{\chi^{(3)}}$:

\begin{align*}
\chi^{(3)}_{xyyy} =3\chi^{(3)}_{xxxy} = 3\chi^{(3)}_{xxyx} = 3\chi^{(3)}_{xyxx} = -3\chi^{(3)}_{yyyx} = -3\chi^{(3)}_{yyxy} = -3\chi^{(3)}_{yxyy}  = -\chi^{(3)}_{yxxx}\equiv \chitrs\\
\end{align*}

where $\chitrs$ quantifies the time-noninvariant terms of  $\boldsymbol{\chi^{(3)}}$ introduced by TRS breaking. These terms are responsible for the rotation~$\theta$ in the TH signal according to the expression 

\begin{align}
\label{eq:rotation_angle_ellipse}
    \tan 2\theta = 2\,\frac{\text{Re}(\chiint\chitrs^*)}{\abs{\chitrs}^2-\abs{\chiint}^2}\approx -2
     \,\frac{\text{Im}(\chitrs)}{\text{Im}(\chiint)}
\end{align}

where $^*$ denotes the complex conjugate. Here, we have assumed $|\chitrs|\ll |\chiint|$ and $\text{Re}(\chiint)=\text{Re}(\chitrs)=0$ at resonance (see Supplementary Information section S1 and S3 for details).

To gain a deeper understanding of the microscopic mechanisms of TRS breaking, we derive analytical expressions for the $\boldsymbol{\chi^{(3)}}$ tensor using perturbative solutions of the SBEs~\cite{Aversa1995,Seith2024,Herrmann2025}. Within this framework, close to the $\pm$K points, a monolayer TMD can be described by a two-band model Hamiltonian~\cite{Herrmann2025}

\begin{align}
    \boldsymbol{h}(\bk) = 
\begin{pmatrix}
     \Delta+\delE  & (-i\kappa_x+\kappa_y\tau)\gamma^*  \\[0.5em]
     (i\kappa_x+\kappa_y\tau)\gamma & -\Delta -\delE
\end{pmatrix}
\label{e27}
\end{align}

where $\Delta$ is the onsite energy ($2\Delta$ is the optical gap), $\gamma$ is the effective hopping,
$\kappa_{x(y)}=a(k_{x(y)}-K_{x(y)})$ is the dimensionless wave vector measured with respect to $\pm$K, $a$ is the TMD lattice constant, and $\tau=\pm1$ is the $\pm$K valley index.
The term $\delE$ is responsible for TRS breaking due to the valley-selective OS and BS effects~\cite{Sie2017,Kim2014}. We evaluate $\delE$ for the elliptical driving electric field given in equation~\eqref{e1} (see details in Supplementary Information section S2) to obtain:

\begin{align}
\delE=\frac{\mathcal{E}^2d^2}{8}
\left(
\frac{1-\tau\sin2\alpha}{2\Delta-\hbar\omega}
+
\frac{1+\tau\sin2\alpha}{2\Delta+\hbar\omega}
\right)
\,.
\label{eqn:delE}
\end{align}

Here, $d$ is the absolute value of the dipole moment which is identical at the $\pm $K valleys. The first and second terms in equation~\eqref{eqn:delE} are the OS and BS shifts, respectively. Close to TH resonance, $3\hbar\omega\approx2\Delta$, we have

\begin{align}
\delE=\frac{3\mathcal{E}^2d^2}{64\Delta}(3-\tau\sin2\alpha)\,.
\end{align}

Next, from perturbative solutions to the SBEs, one can derive an analytical expression for the $\boldsymbol{\chi^{(3)}}$ tensor~\cite{Aversa1995}. We evaluate this expression for the Hamiltonian in equation~\eqref{e27} close to the TH resonance in the leading order in $\delE$ (see Supplementary Information section S3) and we obtain

\begin{align}
\chiint&=\sum_{\tau\pm1}\frac{\mathcal{C}}{2(\Delta+\delE)-3\hbar\omega+{i\hbar}/{T_2}}
=
\frac{2\detuning}{\detuning^2-(\deltagap)^2}\,\mathcal{C}\label{e4}
\,, 
\\[0.5em]
\chitrs&=\sum_{\tau\pm1}\frac{i\tau\mathcal{C}}{2(\Delta+\delE)-3\hbar\omega+{i\hbar}/{T_2}}
=\frac{2i\deltagap}{\detuning^2-(\deltagap)^2}\,\mathcal{C} \,, \label{e5}
\end{align}

with the dephasing time $T_2$~\cite{Herrmann2025}, a constant $\mathcal{C}$ (see Supplementary Information section S3 for details) and the abbreviations

\begin{align}
\detuning&:=2\Delta-3\hbar\omega+\frac{i\hbar}{T_2}+\sum_{\tau=\pm 1} \delE=2\Delta-3\hbar\omega+\frac{i\hbar}{T_2}+\frac{9\mathcal{E}^2d^2}{32\Delta}\,, \\[0.5em]
\deltagap&:=\sum_{\tau=\pm 1} (-\tau) \cdot \delE
=\mathbf{\Delta} E_{ -K,\alpha}-\mathbf{\Delta} E_{ +K,\alpha}
=\frac{3\mathcal{E}^2d^2}{32\Delta}\sin2\alpha\,.
\end{align}

In particular, $\deltagap$ is the difference of the bandgap at $\pm$K caused by the OS and BS effects, which effectively breaks TRS. Finally, we use equations~\eqref{e4} and~\eqref{e5} to evaluate the TH rotation defined in equation~\eqref{eq:rotation_angle_ellipse}: 

\begin{align}
\tan 2\theta=2\,\frac{\text{Re}(\chiint\chitrs^*)}{\abs{\chitrs}^2-\abs{\chiint}^2}
= \frac{-2\deltagap\,\text{Im}\detuning }{|\detuning|^2-|\deltagap|^2}
\,.
\label{e7}
\end{align}

For a small electric field, $\mathcal{E}^2d^2/(\hbar\Delta/T_2)\ll 1$, the TRS breaking~$|\deltagap|\ll |\detuning|$ is small and, thus, the rotation of the TH signal is linear in $\deltagap$ and the intensity:
\begin{align}
\tan 2\theta&=
-\, \frac{2}{1+(2\Delta-3\hbar\omega)^2/(\hbar/T_2)^2}\;\frac{\deltagap}{\hbar/T_2}
\\[0.5em]
&=
-\, \frac{3}{16}\,\sin2\alpha\;\frac{1}{1+(2\Delta-3\hbar\omega)^2/(\hbar/T_2)^2}\;\frac{\mathcal{E}^2d^2}{\hbar\Delta/T_2}\,. \label{e8}
\end{align}

We highlight that the rotation~$\theta$ of the TH signal changes sign when changing the helicity of the driving FB ($\alpha\rightarrow-\alpha$), and that the rotation~$\theta$ is enhanced at the TH resonance $3\hbar\omega=2\Delta$. 

\begin{figure}[ht!]
    \centering
    \includegraphics[width=\linewidth]{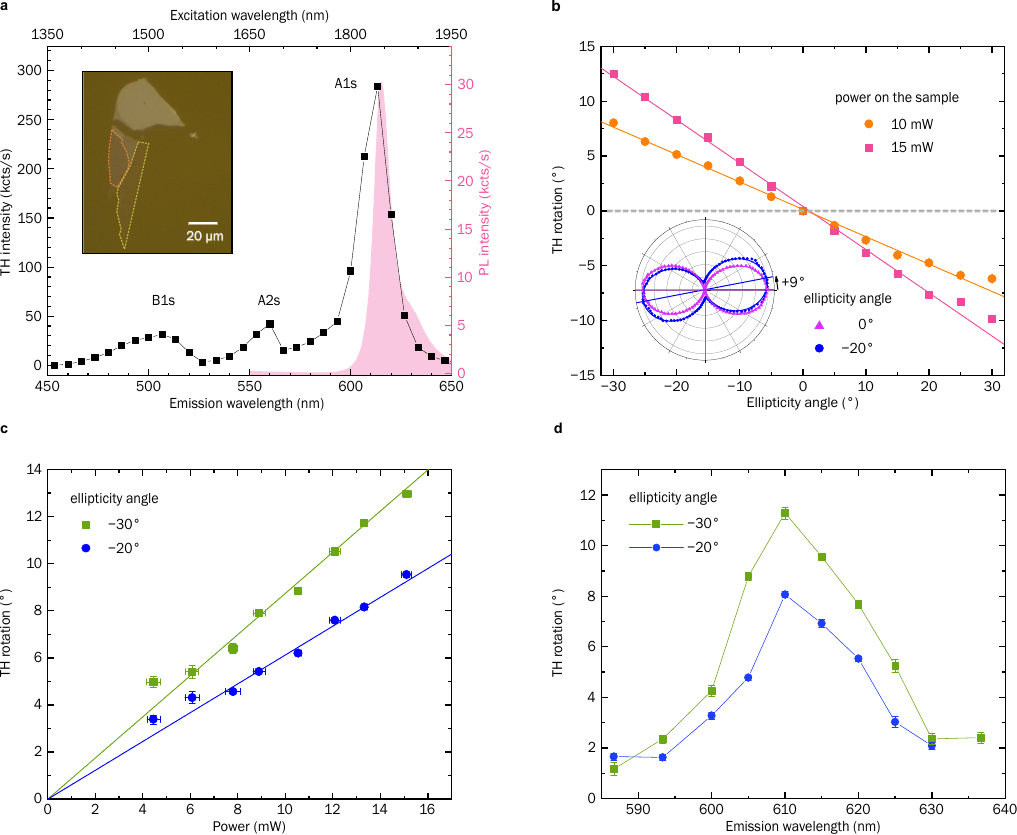}
    \caption{\textbf{PL and TH rotation measurements for different ellipticities and power in monolayer \ce{WS_2}.} \textbf{a} Comparison of the total emitted TH intensity (black squares, left axis) as a function of FB wavelength and the emitted PL (pink line, right axis). \textbf{b} TH rotation angle as a function of the FB ellipticity angle and for an excitation power of \SI{10}{\milli\watt} (orange circles) and \SI{15}{\milli\watt} (pink squares). Solid lines are linear fits to the data. The inset shows the elliptical polarization pattern of the emitted TH for linear (violet triangles) and elliptical (blue circles) input polarization. \textbf{c} Power dependence of the TH rotation for \SI{-20}{\degree} (blue circles) and \SI{-30}{\degree} (green squares). Solid lines are linear fits to the data with a fixed intercept of \SI{0}{\degree} rotation at \SI{0}{\milli\watt}. \textbf{d} Wavelength dependence of the TH rotation angle for an input power of \SI{15}{\milli\watt} for ellipticity angles of \SI{-20}{\degree} (blue circles) and \SI{-30}{\degree} (green squares).}
    \label{fig:FIG3_experimental_results_monolayer}
\end{figure}

We can now move on and compare our theoretical model with experimental results for monolayer \ce{WS_2} (Fig.~\ref{fig:FIG3_experimental_results_monolayer}). The exfoliated \ce{WS_2} sample (see Methods for details) in the inset of Fig.~\ref{fig:FIG3_experimental_results_monolayer}a consists of regions with different number of layers. The monolayer is marked in yellow while the bilayer is marked in orange. Since breaking of TRS is expected to mainly manifest at optical resonances~\cite{Herrmann2025} (see equation~\ref{e8}), we first perform PL and TH wavelength dependence to identify the corresponding FB and TH wavelengths. We find the $A1s$ excitonic resonance at $\SI{615}{nm}$ (Fig.~\ref{fig:FIG3_experimental_results_monolayer}a) in agreement with literature~\cite{Cao2021}. In all following experiments (see Methods for details), the TH rotation angle is measured by rotating a polarizer in front of the detector to obtain a polarization-dependent pattern for the TH signal, as shown in the inset of Fig.~\ref{fig:FIG3_experimental_results_monolayer}b for two exemplary ellipticity angles of \SI{0}{\degree} (linearly polarized) and \SI{-20}{\degree} (elliptically polarized). The polarization-dependent TH patterns are fitted with a $\cos^2$ function, from which we obtain the rotation angle $\theta$ and the error of the numerical fitting. We fix the wavelength of the FB at~\SI{1830}{nm} to work close to the $A1s$ resonance, and perform two different sets of experiments: (1) we fix the input power and tune the ellipticity of the input FB by rotating a quarter-wave plate in front of the sample (Fig.~\ref{fig:FIG3_experimental_results_monolayer}b); (2) we fix the ellipticity and tune the input power (Fig.~\ref{fig:FIG3_experimental_results_monolayer}c). In addition, we scan the wavelength across the $A1s$ resonance with fixed power and ellipticity (Fig.~\ref{fig:FIG3_experimental_results_monolayer}d).

For the first type of experiments (Fig.~\ref{fig:FIG3_experimental_results_monolayer}b) we measure the relative TH rotation angle, namely the difference in the position of the maximum amplitude in detection for elliptical \textit{versus} linear excitation. We observe a clear rotation of the main axis of the polarization ellipse, which can be explained by the new elements of the NLO susceptibility of the $\bar6m'2'$ magnetic point group, introduced by the broken TRS, as discussed above. The TH rotation angle scales linearly with the FB ellipticity: a larger circular component of the FB enhances the effect of TRS breaking. \\ 
We further note that circular THG is forbidden by angular momentum conservation~\cite{Bloembergen1970}, and the total TH intensity scales with the $\cos^2$ of the ellipticity. Thus, since the TH intensity decreases drastically for increasing ellipticity of the FB beam, we investigate the TH rotation only up to an ellipticity angle of \SI{-30}{\degree}. 

\begin{figure}[ht!]
    \centering
    \includegraphics[width=\linewidth]{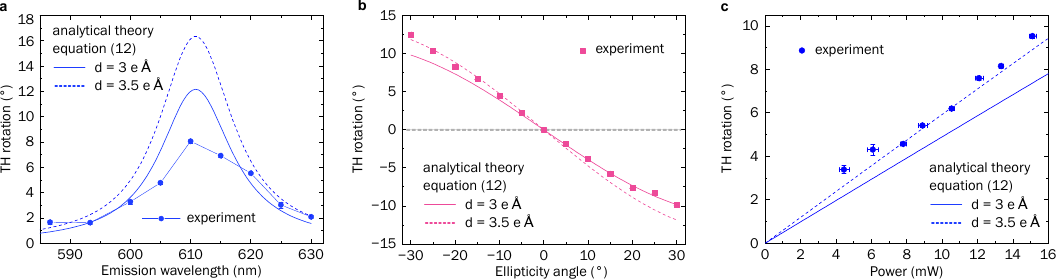}
    \caption{\textbf{Comparison of analytical model and experimental results for monolayer \ce{WS_2}.} \textbf{a} Wavelength-dependent TH rotation for fixed ellipticity of \SI{-20}{\degree} and input power of \SI{15}{\milli\watt} (blue circles in Fig.~\ref{fig:FIG3_experimental_results_monolayer}d). \textbf{b} Ellipticity-dependent TH rotation for a fixed input power of \SI{15}{\milli\watt} (pink squares in Fig.~\ref{fig:FIG3_experimental_results_monolayer}b). \textbf{c} Power-dependent TH rotation for a fixed ellipticity of \SI{-20}{\degree} (blue circles in Fig.~\ref{fig:FIG3_experimental_results_monolayer}c).
    For the analytical calculations, we employ $T_2\eqt 28\,\text{fs}$,  $\Delta\eqt$ \SI{1.05}{\electronvolt}  and two values of the dipole element: $d\eqt\SI{3}{\eangstrom}$ (solid lines) and $d\eqt\SI{3.5}{\eangstrom}$ (dashed lines).
    }
    \label{fig:FIG4_theory_monolayer}
\end{figure}

In the second set of experiments, we fix the ellipticity angle to \SI{-30}{\degree} and \SI{-20}{\degree} and scan the input power in the range from \SI{4.5}{\milli\watt} to \SI{15}{\milli\watt}. The results are shown in Fig.~\ref{fig:FIG3_experimental_results_monolayer}c, where we observe a close to linear dependence of the TH rotation angle with respect to the input power, as expected for TRS breaking due to OS and BS shifts~\cite{Herrmann2025}. We highlight that in the fitting of Fig.~\ref{fig:FIG3_experimental_results_monolayer}c we fixed the intercept to zero, because the TH rotation must be zero for an unperturbed sample. Finally, we measure the wavelength dependence of the TH rotation angle close to the exciton resonance for two different ellipticity angles of \SI{-20}{\degree} and \SI{-30}{\degree} and a fixed input power of \SI{15}{\milli\watt} on the sample (Fig.~\ref{fig:FIG3_experimental_results_monolayer}d). 
We observe the largest TH rotation for excitation at the $A1s$ resonance, while the rotation angle decreases for off-resonant wavelengths. 

Next, we quantitatively compare the experimental results to the analytical formula~\eqref{e8} in Fig.~\ref{fig:FIG4_theory_monolayer}. 
According to equation~\eqref{e8},  the TH rotation~$\theta$ depends on the ellipticity angle~$\alpha$, frequency~$\omega$ and field strength~$\mathcal{E}$ of the fundamental beam, which are known. 
Furthermore, $\theta$  also depends on material parameters: optical gap~$2\Delta$, dipole element~$d$, and dephasing time~$T_2$.
To determine $2\Delta$ and $T_2$, we focus on the peak of $\theta(\omega)$ in equation~\eqref{e8} at the TH resonance, where  $3\omega\eqt 2\Delta$.  
From the peak position \SI{615}{\nano\meter} in the wavelength scan (Fig.~\ref{fig:FIG4_theory_monolayer}a), we extract~$2\Delta\eqt \SI{2.1}{\electronvolt}$ and from the linewidth of the peak, we obtain $T_2\eqt \SI{28}{\femto\second}$. 
We note that the theoretical model only includes dephasing mechanisms such as electron-electron and electron-phonon scattering, leading to a homogeneous broadening. However, the experiments are sensitive to any homogeneous and inhomogeneous broadening mechanism, including space-local band gap modulations caused by defects and strain~\cite{vandeGroep2023}.
Our extracted value $T_2\eqt28$\,fs effectively includes all of these broadening mechanisms. 

After determining~$\Delta$ and $T_2$, we report the TH rotation $\theta$ from equation~\eqref{e8} for two values of the dipole element, $d\eqt$\SI{3}{\eangstrom} and $d\eqt$\SI{3.5}{\eangstrom}.
The theoretical and experimental results show good agreement in their wavelength dependence, with peak heights matching within a factor of two (Fig.~\ref{fig:FIG4_theory_monolayer}a).
%
For the ellipticity dependence (Fig.~\ref{fig:FIG4_theory_monolayer}b), the analytical SBE model gives a nonlinear curve which describes with excellent accuracy the experimental results. 
The deviation from the linear dependence, which we use for simplicity in Fig.~\ref{fig:FIG3_experimental_results_monolayer}b, can be understood from the $\sin 2\alpha$ dependence in equation~\eqref{e8}, assuming small $\theta$ such that $\tan\theta\approx\theta\propto \sin 2\alpha$.
Similarly, we find the linear increase of $\theta$ with the fundamental power both in experiment and in the analytical model (\textit{via} $\mathcal{E}^2$ in equation \eqref{e8}, Fig.~\ref{fig:FIG4_theory_monolayer}c). 
As previously discussed, the proportionality $\theta\propto \mathcal{E}^2 \sin 2\alpha$ is a direct consequence of the OS and BS shifts.


\subsection{TH Kerr rotation in centrosymmetric bilayer WS\textsubscript{2}}

Having demonstrated the potential of TH Kerr rotation in non-centrosymmetric monolayer \ce{WS_2}, we now demonstrate the possibility to probe broken TRS in centrosymmetric bilayer \ce{WS_2}. 

\begin{figure}
    \centering
    \includegraphics[width=\linewidth]{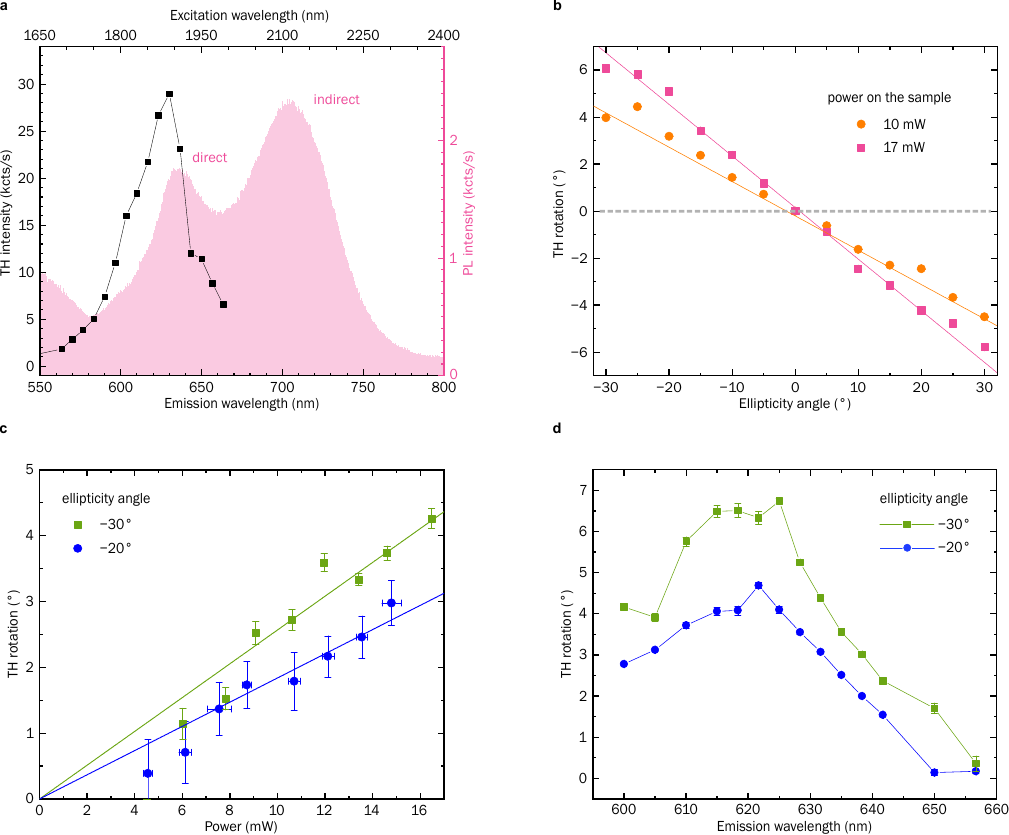}
    \caption{\textbf{PL and TH rotation measurements for different ellipticity angles and power values in bilayer \ce{WS_2}.} \textbf{a} Comparison of the total emitted TH intensity (black squares, left axis) and the emitted PL (pink line, right axis) as a function of the excitation wavelength. \textbf{b} Rotation of the emitted TH dependent on the ellipticity angle and for an excitation power of \SI{10}{\milli\watt} (orange circles) and \SI{17}{\milli\watt} (pink squares). Solid lines are linear fits to the data. \textbf{c} Power dependence of the TH rotation for \SI{-20}{\degree} (blue circles) and \SI{-30}{\degree} (green squares). Solid lines are linear fits to the data with a fixed intercept of \SI{0}{\degree} rotation at \SI{0}{\milli\watt}. \textbf{d} Wavelength dependence of the TH rotation for an input power of \SI{15}{\milli\watt} for ellipticity angles of \SI{-20}{\degree} (blue circles) and \SI{-30}{\degree} (green squares).}
    \label{fig:FIG5_experimental_results_bilayer}
\end{figure}

Breaking of TRS in TMD bilayers follows a similar principle as in the monolayer case (Fig.~\ref{fig:FIG2_schematics_mono_bi}g-l): an elliptical input pulse leads to alternating gap opening (in the $+$K valley of the upper layer, in the $-$K valley of the bottom layer) due to spin-valley-layer locking~\cite{Jones2014}, see Fig.~\ref{fig:FIG2_schematics_mono_bi}l. When TRS is broken, the valleys remain energetically degenerate, but they are no longer spin-degenerate. This can be understood by considering the energy-spin relation of a system where SIS is preserved ($E_{\uparrow} (+\bk) = E_{\uparrow} (-\bk) $) and TRS is broken ($E_{\uparrow} (+\bk) \neq E_{\downarrow} (-\bk) $). Looking at the magnetic point groups, bilayer TMDs belong to $\bar3m$ when TRS is preserved, and to $\bar3m'$ when TRS is broken. These point groups have the same in-plane components of the $\boldsymbol{\chi^{(3)}}$ susceptibility as the $\bar6m2$ (when TRS is preserved) and $\bar6m'2'$ (when TRS is broken) groups of monolayer TMDs. Thus, for in-plane excitation of the sample, we can apply exactly the same considerations done for monolayer TMDs also to the case of centrosymmetric bilayer TMDs. 

To demonstrate that this is indeed the case, we perform the same set of experiments discussed in Fig.~\ref{fig:FIG3_experimental_results_monolayer} also for a centrosymmetric bilayer \ce{WS_2} crystal (see Fig.~\ref{fig:FIG5_experimental_results_bilayer}). 

We start with PL and TH wavelength dependence to determine the $\pm $K direct transitions of the \ce{WS_2} bilayer (Fig.~\ref{fig:FIG5_experimental_results_bilayer}a). The PL signal shows two maxima, which we can assign to the momentum direct and phonon-assisted momentum indirect transitions~\cite{Echeverry2016,doi:10.1021/acs.jpcc.9b08602} at \SI{635}{\nano\meter} and \SI{705}{\nano\meter}, respectively. We note that momentum indirect transitions can not be probed by coherent and parametric harmonic generation, as they require exchange of momentum with \textit{e.g.} defects or phonons. For an emission wavelength of \SI{625}{\nano\meter} (FB at \SI{1875}{\nano\meter}) we observe the maximum TH and assign the difference between PL and TH measurements to the presence of trions~\cite{Mak2013}. Thus, in our TH experiments, we will study only the momentum direct $\pm $K transitions at a FB wavelength of \SI{1875}{\nano\meter}. 

In analogy with the monolayer case, the ellipticity dependence (Fig.~\ref{fig:FIG5_experimental_results_bilayer}b) shows a close to linear dependence of the TH rotation angle \textit{versus} the ellipticity of the FB. As discussed, the in-plane elements of the third order NLO susceptibility are identical in monolayer and bilayer TMDs when TRS is preserved/broken, and, thus, the TH rotation in the case of bilayer \ce{WS_2} follows the same rules as discussed in the previous paragraph. However, we notice that, for a given ellipticity of the input FB, the TH rotation is smaller in the bilayer compared to the monolayer case. For example, we observe a rotation of \SI{8}{\degree} at \SI{-20}{\degree} for the monolayer sample and of \SI{4}{\degree} for the bilayer sample, both at a power of \SI{10}{\milli\watt}. According to equation~\eqref{eq:rotation_angle_ellipse}, the rotation angle only depends on the ratio ${\text{Im}(\chitrs)}/{\text{Im}(\chiint)}$. Thus, this result could be explained either by a larger intrinsic $\chiint$ or by a smaller $\chitrs$. The intrinsic $\chiint$ is, by definition, independent of the sample thickness. In addition, we observe a higher resonant TH intensity in the monolayer compared to the bilayer case, indicating that the resonant intrinsic $\chiint$ is weaker in the latter. Thus, we conclude that the smaller TH rotation angle observed in bilayer compared to monolayer sample is a consequence of a weaker $\chitrs$ from broken TRS. We predict and infer that this different intensity in the $\chitrs$ between monolayer and bilayer TMDs directly arises from the different mechanisms that underlie broken TRS in the two systems, namely valley-exclusive bandgap opening in monolayer TMDs \textit{versus} valley-symmetric but spin-selective bandgap opening in bilayer TMDs. Thus, further studies in this direction will shed light on the interplay between TRS and the coupling of valley, spin and layer in TMDs.

Next, we study the power dependence of the TH rotation angle (Fig.~\ref{fig:FIG5_experimental_results_bilayer}c) for different values of the ellipticity angle (\SI{-20}{\degree} and \SI{-30}{\degree} in blue circles and green squares, respectively). We observe a close to linear dependence, which is again a clear indication that also in bilayer TMDs broken TRS, for our experimental conditions, arises from all-optical bandgap modulation induced by the OS and BS shifts. However, we highlight again that in bilayer TMDs these two coherent effects are valley-symmetric, rather than valley-exclusive as for monolayer TMDs~\cite{Sie2017,Kim2014}.

Finally, we measure the wavelength dependence of the TH rotation angle for two different ellipticity angles of \SI{-20}{\degree} and \SI{-30}{\degree} and input power of \SI{15}{\milli\watt} (Fig.~\ref{fig:FIG5_experimental_results_bilayer}d). Similar to the monolayer case, we observe the largest TH rotation for excitation at \SI{625}{\nano\meter}, \textit{i.e.} at resonance with the momentum direct $\pm$K optical transitions. 

\section{Conclusion}

We have demonstrated an all-optical method to probe broken TRS in both centrosymmetric and non-centrosymmetric systems. The approach is based on the measurement of the TH Kerr rotation, namely a rotation of the TH polarization angle with respect to the polarization of the input FB. We realize and test TH Kerr rotation in two exemplary crystals: monolayer \ce{WS_2} for the non-centrosymmetric case and bilayer \ce{WS_2} for the centrosymmetric case. In both systems, broken TRS has fundamental implications for the understanding of the light-modulated band structure as well as for technological applications. For monolayers, broken TRS induces a valley imbalance, which could be the source of information for valleytronic operations. In bilayer TMDs, broken TRS interacts with the spin-valley-layer locking, that is unique of this type of layered van der Waals structures. For both samples, we show that broken TRS is induced by all-optical and ultrafast OS and BS effects, and the effect is probed by a sizeable TH rotation angle up to \SI{12.5}{\degree} under our experimental conditions. These results represent an important step towards the development of new all-optical diagnostic techniques to probe broken TRS in any system, regardless of SIS. In addition, this work provides a viable approach for the realization of all-optical and ultrafast valleytronic devices. 

\section{Methods}

\subsection{Sample preparation and characterization}

The mono- and multilayer \ce{WS_2} sample was mechanically exfoliated onto polydimethylsiloxane (PDMS) from a bulk crystal (HQ graphene), and then transferred onto a fused silica substrate. The different thicknesses (mono- and bilayer) were confirmed by PL and contrast measurements. For the PL measurements, we used a \SI{532}{\nano\meter} diode laser (Cobolt 08-DPL \SI{532}{\nano\meter}) as excitation source, which we focus onto the sample with a 50x objective (M Plan Apo 50x, Mitutoyo). The sample was mounted on a motorized 3-axis stage under a home-built microscope. The emitted PL was collected in reflection geometry and guided to the spectrometer (Horiba iHR550) after proper filtering with a longpass dichroic mirror (LPD02-532RU-25, Semrock) and a notch filter (FL532-3, Thorlabs).

\subsection{Polarization-resolved THG}

For all the THG experiments, we use a home-built multi-photon microscope in transmission geometry. An optical parametric oscillator (Levante IR fs from APE), pumped by the output of an Yb-doped mode-locked laser (FLINT FL2-12, LIGHT CONVERSION) with a repetition rate of \SI{76}{\mega\hertz} and a pulse length of $\thicksim \SI{100}{\femto\second}$, generates the FB, which is tunable in the range \SI{1300}{\nano\meter} to \SI{2000}{\nano\meter}. To precisely control the average power of the FB, we use a combination of a fixed Glan-Thompson polarizer (GTH10M, Thorlabs) and a wire grid polarizer (WP25M-UB, Thorlabs), which is mounted on a motorized rotation stage (RSP05/M, Thorlabs). Subsequently, the ellipticity is controlled \textit{via} a quarter-wave plate (RSU 2.4.15, B. Halle) mounted on another motorized rotation stage (RSP05/M, Thorlabs). The FB is focused onto the sample by a 40x mirror objective (LMM-40x-UVV) with a focal spot radius of \SI{1.85}{\micro\meter}. The transmitted FB as well as the emitted TH are collimated by a lens (C330TMD, Thorlabs). To obtain polarization-resolved results, we place another wire grid polarizer (WP25M-UB, Thorlabs) after the collimation and before the detector. We block the transmitted FB by two heat absorbing filters (Isuzu ISK171) as well as a shortpass filter (FESH0700, Thorlabs). Furthermore, bandpass filters for \SI{610}{\nano\meter} (FBH610-10, Thorlabs) and \SI{625}{\nano\meter} (87791, Edmund Optics) can be added. The TH signal is then detected by a silicon avalanche photodiode (APD440A, Thorlabs) and filtered by a lock-in amplifier (MFLI, Zurich Instruments) in combination with an optical chopper (MC2000B, Thorlabs) that modulates the FB at \SI{971}{\hertz}. 

\subsection{Analytical methods}

For the analytical calculation of polarization-resolved THG, we employ an analytical expression of the third-order NLO susceptibility developed in Ref.~\cite{Aversa1995}. Details are given in the Supplementary Information section S2 and S3.

\section{Author contribution}

F.F., P.H. and G.S. conceived the work. F.F. and P.H. performed the measurements and analyzed the data. F.F., P.H. and G.S. interpreted experimental results. S.K. fabricated the sample. S.S.S. and J.W. developed the analytical model. The manuscript was written with contributions from all co-authors.

\section{Acknowledgements}
S.S.S. thanks Bashab Dey for helpful discussion. This work was funded by the German Research Foundation (DFG) (CRC 1375 NOA), project number 398816777 (subproject C4); the International Research Training Group (IRTG) 2675 ‘Meta-Active’, project number 437527638 (subproject A4); and the Federal Ministry for Education and Research (BMBF) project number 16KIS1792 SiNNER. J.W. acknowledges the DFG for funding via the Emmy Noether Programme (project number 503985532), CRC 1277 (project number 314695032, subproject A03) and RTG 2905 (project number 502572516).

\bibliographystyle{naturemag}

\end{document}



\section{Supplementary Information}
Direct measurement of broken time-reversal symmetry in centrosymmetric and non-centrosymmetric atomically thin crystals with nonlinear Kerr rotation

\section{Author list}
Florentine Friedrich$^1$, Paul Herrmann$^1$, Shridhar Sanjay Shanbhag$^2$, Sebastian Klimmer$^{1,3}$, Jan Wilhelm$^{2}$, and {Giancarlo} Soavi$^{1,4,\star}$

\section{Affiliations}
\noindent
$^1$Institute of Solid State Physics, Friedrich Schiller University Jena, Helmholtzweg 5, 07743 Jena, Germany
\newline
$^2$Institute of Theoretical Physics and Regensburg Center for Ultrafast Nanoscopy (RUN), University of Regensburg, Universitätsstrasse 31, 93053 Regensburg, Germany
\newline
$^3$ARC Centre of Excellence for Transformative Meta-Optical Systems, Department of Electronic Materials Engineering, Research School of Physics, The Australian National University, Canberra ACT 2601, Australia
\newline
$^4$Abbe Center of Photonics, Friedrich Schiller University Jena, Albert-Einstein-Straße 6, 07745 Jena, Germany
\newline
$^{\star}$ giancarlo.soavi@uni-jena.de

\maketitle
\tableofcontents

\newpage

\section{S1 Elliptical Third Harmonic Generation -- classical derivation based on magnetic point groups}

\noindent
We start by comparing the $\boldsymbol{\chi^{(3)}}$ elements of the magnetic point groups \textcolor{cyan}{$\overline{6}m2$} and \textcolor{cyan}{$\bar3m$} (mono- and bilayer TMD with preserved TRS), and \textcolor{blue}{$\overline{6}m'2'$} and \textcolor{blue}{$\bar3m'$} (mono- and bilayer TMD with broken TRS). The NLO tensors for magnetic point groups are available in Refs.~\cite{Gallego:lk5043, MTensor}. We assume normal incidence on the TMD plane and thus neglect the z components of the electric field. 
\noindent
For the \textcolor{cyan}{$\overline{6}m2$} and \textcolor{cyan}{$\bar3m$} point groups, the non-zero in-plane elements of the $\boldsymbol{\chi^{(3)}}$ tensor are identical:

\begin{align}
    \chi^{(3)}_{xxxx} &= \chi^{(3)}_{yyyy} = 3\chi^{(3)}_{xxyy} =3 \chi^{(3)}_{xyyx} =3 \chi^{(3)}_{xyxy} \equiv \chiint  \,.
\end{align}

In addition to these terms, we have the following non-zero elements of the NLO susceptbility for the \textcolor{blue}{$\overline{6}m'2'$} and \textcolor{blue}{$\bar3m'$} magnetic point groups: 

\begin{align}
\chi^{(3)}_{xyyy} =3\chi^{(3)}_{xxxy} = 3\chi^{(3)}_{xxyx} = 3\chi^{(3)}_{xyxx} = -3\chi^{(3)}_{yyyx} = -3\chi^{(3)}_{yyxy} = -3\chi^{(3)}_{yxyy}  = -\chi^{(3)}_{yxxx}\equiv \chitrs \,,
\end{align}
where $\chitrs$ quantifies the $\boldsymbol{\chi^{(3)}}$ elements introduced by TRS breaking. We can thus write the third harmonic polarization for an arbitrary input and for all the aforementioned magnetic point groups as: 

\begin{equation}
    \boldsymbol{P}(3\omega) = \varepsilon_0
    \begin{pmatrix}
    \textcolor{cyan}{\chi^{(3)}_{xxxx}} & \textcolor{blue}{\chi^{(3)}_{xyyy}} & \textcolor{cyan}{\left(\chi^{(3)}_{xxyy}+\chi^{(3)}_{xyyx}+\chi^{(3)}_{xyxy}\right)} & \textcolor{blue}{\left(\chi^{(3)}_{xxxy} + \chi^{(3)}_{xxyx} + \chi^{(3)}_{xyxx}\right)} \\
    \textcolor{blue}{\chi^{(3)}_{yxxx}} & \textcolor{cyan}{\chi^{(3)}_{yyyy}} & \textcolor{blue}{\left(\chi^{(3)}_{yxyy} + \chi^{(3)}_{yyyx} + \chi^{(3)}_{yyxy}\right)} & \textcolor{cyan}{\left(\chi^{(3)}_{yxxy} + \chi^{(3)}_{yxyx} + \chi^{(3)}_{yyxx}\right)} 
    \end{pmatrix}
    \begin{pmatrix}
        E_x^3 \\
        E_y^3 \\
        E_xE_y^2 \\
        E_y^2 E_x
    \end{pmatrix} \,,
\end{equation}
\noindent
where \textcolor{cyan}{the} the blue elements are non-zero only in the \textcolor{blue}{$\overline{6}m'2'$} and \textcolor{blue}{$\bar3m'$} magnetic point groups. Furthermore, the following two expressions

\begin{align}
\chi^{(3)}_{xxxy} + \chi^{(3)}_{xxyx} + \chi^{(3)}_{xyxx}  &= (\chi^{(3)}_{yyxy} + \chi^{(3)}_{yxyy} + \chi^{(3)}_{xyyy}) + \chi^{(3)}_{xxyx} + \chi^{(3)}_{xyxx} \nonumber \\
                    &= (-\chi^{(3)}_{xxyx} - \chi^{(3)}_{xyxx} - \chi^{(3)}_{yxxx}) + \chi^{(3)}_{xxyx} + \chi^{(3)}_{xyxx} \nonumber \\
                    &=-\chi^{(3)}_{yxxx} \nonumber \\
                    &=\chi^{(3)}_{xyyy}
\end{align}
 and 
\begin{align}
\chi^{(3)}_{yxyy} + \chi^{(3)}_{yyyx} + \chi^{(3)}_{yyxy} &= \chi^{(3)}_{yxyy} - (\chi^{(3)}_{yyxy} + \chi^{(3)}_{yxyy} + \chi^{(3)}_{xyyy}) + \chi^{(3)}_{yyxy} \nonumber \\
                    &= \chi^{(3)}_{yxyy} - \chi^{(3)}_{yyxy} - \chi^{(3)}_{yxyy} - \chi^{(3)}_{xyyy} + \chi^{(3)}_{yyxy} \nonumber \\
                    &=-\chi^{(3)}_{xyyy}
\end{align}
\noindent

allow us to rewrite the TH polarization as:

\begin{align}
    \boldsymbol{P}(3\omega) &= \varepsilon_0
    \begin{pmatrix}
    \chi_\mathrm{int} & \chi_\mathrm{TRS} & \chi_\mathrm{int} & \chi_\mathrm{TRS} \\
    -\chi_\mathrm{TRS} & \chi_\mathrm{int} & -\chi_\mathrm{TRS} & \chi_\mathrm{int}
    \end{pmatrix}
    \begin{pmatrix}
        E_x^3 \\
        E_y^3 \\
        E_xE_y^2 \\
        E_x^2 E_y
    \end{pmatrix} \nonumber \\
    &=  \varepsilon_0
    \begin{pmatrix}
    \chi_\mathrm{int} (E_x^3 + E_xE_y^2) +\chi_\mathrm{TRS} (E_y^3 + E_x^2 E_y)\\
    \chi_\mathrm{int} (E_y^3 + E_x^2 E_y) -\chi_\mathrm{TRS} (E_x^3 + E_xE_y^2)
    \end{pmatrix} .
\end{align}

Next, we rewrite the FB electromagnetic field in the case of elliptical polarization: 

\begin{align}
    \boldsymbol{E}_\mathrm{IN} = \mathcal{E}\,\exp{i\frac{\pi}{4}}
    \begin{pmatrix}
        1 & 0 \\
        0 & i
    \end{pmatrix} 
    \cdot
    \begin{pmatrix}
    \cos\alpha \\
    \sin\alpha
    \end{pmatrix} = \mathcal{E}\, \exp{ i\frac{\pi}{4}}  \begin{pmatrix}
    \cos\alpha \\
    i \sin\alpha
    \end{pmatrix} \,,\label{e7f}
\end{align}
where $\mathcal{E}$ is the amplitude of the driving field. 
\noindent

Without loss of generality, we choose a horizontal fast axis of the QWP and neglect the constant phase factor $\exp{i\frac{\pi}{4}}$~\cite{Hecht2002}. Note that the pre-factor $\exp{- i\frac{\pi}{4}}$ appears only if one defines the phase delays in a symmetric way: $\varphi_x = - \varphi_y =  \frac{\pi}{4}$. This is done for instance in Ref.~\cite{Hecht2002}, and not in Ref.~\cite{Fowles1989}. Based on this, we can rewrite the TH polarization in the case of elliptical FB input as:

\begin{align}
    \boldsymbol{P}(3\omega) &= \varepsilon_0\, \mathcal{E}^3\begin{pmatrix}
    \chi_\mathrm{int} (\cos\alpha \cos2\alpha) + i \chi_\mathrm{TRS} \left (\sin\alpha \cos2\alpha \right )\\
    i \chi_\mathrm{int} (\sin\alpha \cos2\alpha) - \chi_\mathrm{TRS} (\cos\alpha \cos2\alpha)
    \end{pmatrix} \nonumber \\
    &= \varepsilon_0\, \mathcal{E}^3 \cos2\alpha\,\
    \begin{pmatrix}
    \chi_\mathrm{int} \cos\alpha + i \chi_\mathrm{TRS} \sin\alpha\\
    i \chi_\mathrm{int} \sin\alpha - \chi_\mathrm{TRS} \cos\alpha
    \end{pmatrix} .
\end{align}
\noindent 

For $\alpha = \pm \frac{\pi}{4}$ (circularly polarized light), we obtain $\boldsymbol{P}(3\omega) = 0$ regardless of the values of $\chi_\mathrm{int}$ and $\chi_\mathrm{TRS}$, as expected from the conservation of angular momentum in THG for crystals with 3-fold rotational symmetry~\cite{Bloembergen1970}. Otherwise, an elliptical FB input generates an elliptical TH polarization. Based on the Stokes parameters, we can define the rotation angle $\theta$ of TH elliptical polarization as:

\begin{align}
    \tan 2\theta = \frac{S_2}{S_1}\,.
\end{align}
\noindent If we write the complex NLO susceptibilities with their real and imaginary parts as \mbox{$\chi_\mathrm{int}\coloneqq a + i \tilde{a}$} and $\chi_\mathrm{TRS} \coloneqq b + i \tilde{b}$, we obtain for the TH polarization:

\begin{align}
    \boldsymbol{P}(3\omega)= \varepsilon_0\, \mathcal{E}^3 \cos2\alpha\,\ \begin{pmatrix}
    \left(a + i \tilde{a}\right) \cos\alpha + i\left(b + i \tilde{b}\right) \sin\alpha\\
    i  \left(a + i \tilde{a}\right) \sin\alpha - \left(b + i \tilde{b}\right) \cos\alpha
    \end{pmatrix} \coloneqq 
        \begin{pmatrix}
            p_x\\
            p_y
        \end{pmatrix} + i 
        \begin{pmatrix}
            q_x\\
            q_y
        \end{pmatrix}\,,
\end{align}
\noindent 

from which we can calculate the Stokes parameters:

\begin{align}
S_1 &\propto 
\abs{P_x}^2 - \abs{P_y}^2 \nonumber\\
&= p_x^2 + q_x^2 - p_y^2 -q_y^2 \nonumber \\
&\propto \cos^2 2\alpha \left[ (a^2 + \tilde{a}^2 -  b^2 - \tilde{b}^2) \cos^2 \alpha + (\tilde{b}^2  +  b^2 - \tilde{a}^2 -a^2) \sin^2\alpha \right] \nonumber \\
&= \cos^3 2\alpha \left(\abs{\chi_\mathrm{int}}^2 - \abs{\chi_\text{TRS}}^2 \right)
\end{align}
\noindent

and  
\begin{equation}
    S_2 \propto 2 \, (p_x  p_y + q_x q_y)\propto - 2 \,  \cos^3 2\alpha \left(ab+\tilde{a} \tilde{b}\right) .
\end{equation}
\noindent

From this we obtain a rotation angle of the TH elliptical polarization of

\begin{align}
    \tan 2\theta = \dfrac{S_2}{S_1} = \dfrac{- 2 \,  \cos^3 2\alpha \left(ab+\tilde{a} \tilde{b}\right)}{\cos^3 2\alpha \left(\abs{\chi_\mathrm{int}}^2 - \abs{\chi_\mathrm{TRS}}^2 \right)}
\end{align}\
\noindent 

which simplifies to the same expression of elliptical SHG~\cite{Herrmann2023}:

\begin{align}
    \tan 2\theta \approx -2 \, \dfrac{ab + \tilde{a}\tilde{b}}{\abs{\chi_\mathrm{int}}^2} \approx -2
     \,\frac{\text{Im}(\chitrs)}{\text{Im}(\chiint)}
\end{align}

in the limit $\abs{\chi_\mathrm{int}} \gg \abs{{\chi}_\mathrm{TRS}}$ and considering that at optical resonances the NLO susceptibility is purely imaginary ($\text{Re}(\chiint)=\text{Re}(\chitrs)=0$).

\section{S2 Optical Stark and Bloch-Siegert shifts with elliptical polarization}
An off-resonant electromagnetic field radiating on a material induces an energy shift that can be obtained from time-dependent perturbation theory. Here, we follow the semi-classical derivation used in Ref.~\cite{Sie2017}.
We express the incoming elliptically polarized light~\eqref{e7f} of ellipticity angle $\alpha$ and frequency $\omega$ in the time domain as
\begin{align}
\boldsymbol{{\mathcal{E}}}(t)&=\mathcal{E}
\left( 
\begin{array}{c}
\cos\alpha\cos \omega t \\
\sin\alpha\sin \omega t
\end{array}
\right) =\frac{\mathcal{E}}{2}\left[
e^{ i\omega t}
\left(
\begin{array}{c}
\cos\alpha\\
-i\sin\alpha
\end{array}
\right)
+e^{-i\omega t}
\left(
\begin{array}{c}
\cos\alpha\\
i\sin\alpha
\end{array}
\right)
\right]\,.
\end{align}
The perturbation Hamiltonian that is induced by $\boldsymbol{\mathcal{E}}$ is~\cite{Sie2017},
\begin{align}
H_{vc}(t)&=\boldsymbol{d_{cv}}(\tau)\cdot\boldsymbol{\mathcal{E}}(t)\nonumber\\
&=\frac{\mathcal{E}}{2}\left[
e^{ i\omega t}\left( d^{x}_{cv}(\tau)\cos\alpha- i d^{y}_{cv}(\tau)\sin \alpha\right)+
e^{-i\omega t}\left( d^{x}_{cv}(\tau)\cos\alpha+ i d^{y}_{cv}(\tau)\sin \alpha\right)
\right]\nonumber\\[0.2em]
&=\frac{\mathcal{E}d_0}{2}\left[
e^{ i\omega t}\left(\cos \alpha + \tau \sin \alpha \right)+
e^{-i\omega t}\left(\cos \alpha - \tau \sin \alpha \right)
\right]\,.
\end{align}
Here, ${d}_{cv}^x(\tau)\eqt d_0$ and $ d_{cv}^y(\tau)\eqt i\tau d_0$ are the dipole matrix elements~\cite{Herrmann2025} between the valence band maximum and the conduction band minimum at the $\pm$K points of a TMD, $\pm$K$=\frac{2\pi}{3}(\sqrt{3},\tau)$, where $\tau=\pm1$. 

The induced energy shift of the valence band can be obtained through the standard time-dependent perturbation theory as done in~\cite{Sie2017},
\begin{align}
\Delta E_v(t)&=H_{vc}^*(t)e^{-i\omega_0 t}\frac{1}{i\hbar}\int_0^t H_{vc}(t')e^{i\omega_0 t'}dt'\nonumber \\
&=\frac{\mathcal{E}d_0^*}{2i\hbar}\left[
e^{ i\omega t}\left(\cos \alpha - \tau \sin \alpha \right)+
e^{-i\omega t}\left(\cos \alpha + \tau \sin \alpha \right)
\right]
e^{-i\omega_0 t}\nonumber\\
&\qquad\int_0^t
\frac{\mathcal{E}d_0}{2}\left[
e^{ i\omega t'}\left(\cos \alpha + \tau \sin \alpha \right)+
e^{-i\omega t'}\left(\cos \alpha - \tau \sin \alpha \right)
\right]
e^{i\omega_0 t'}dt'\nonumber\\
&=\frac{\mathcal{E}^2\abs{d_0}^2}{4i\hbar}\left[
e^{-i(\omega_0-\omega) t}\left(\cos \alpha - \tau \sin \alpha \right)+
e^{-i(\omega_0+\omega) t}\left(\cos \alpha + \tau \sin \alpha \right)
\right]\nonumber\\
&\qquad\left[
\left(\frac{e^{i(\omega_0+\omega) t}-1}{i(\omega_0+\omega)}\right)(\cos \alpha + \tau \sin \alpha)+
\left(\frac{e^{i(\omega_0-\omega) t}-1}{i(\omega_0-\omega)}\right)(\cos \alpha - \tau \sin \alpha)
\right]\nonumber\\
&=\frac{\mathcal{E}^2\abs{d_0}^2}{4i\hbar}
\Bigg[
\left(\frac{e^{2i\omega t}-e^{-i(\omega_0-\omega) t}}{i(\omega_0+\omega)}+\frac{e^{-2i\omega t}-e^{-i(\omega_0+\omega) t}}{i(\omega_0-\omega)}\right)
(\cos^2\alpha - \sin^2 \alpha)\nonumber\\
&\qquad+\left(\frac{1-e^{-i(\omega_0-\omega) t}}{i(\omega_0-\omega)}\right)
(1-2\tau\sin\alpha\cos\alpha) +\left(\frac{1-e^{-i(\omega_0+\omega) t}}{i(\omega_0+\omega)}\right)
(1+2\tau\sin\alpha\cos\alpha)\Bigg]\,,
\end{align}
where $\omega_0\eqt \varepsilon_{cv}/\hbar$ is the frequency associated with the band gap~$\varepsilon_{cv}$ at $\pm$K.
The energy shift $\Delta E_v(t)$ saturates at its mean value, which can be obtained by averaging over time,
\begin{align}
\overline{\Delta E_v}&=\lim_{T\to\infty}\frac{1}{T}\int_0^T\Delta E_v(t)dt=-\frac{\mathcal{E}^2\abs{d_0}^2}{4}
\left(
\frac{1+\tau\sin2\alpha}{\varepsilon_{cv}+\hbar \omega}+
\frac{1-\tau\sin2\alpha}{\varepsilon_{cv}-\hbar\omega}
\right) =: -\delE \,.
\end{align}
For the conduction band, the energy is the same but with an opposite sign,
\begin{align}
\overline{\Delta E_c}=-\overline{\Delta E_v}= \delE\,,
\end{align}
The gap opening $2\delE$ depends on the valley index~$\tau$ and thus is  responsible for breaking the TRS. For further simplification, we employ 
the absolute value of the dipole moment  $d\eqt \sqrt{|d^x_{cv}|^2+|d^y_{cv}|^2}\eqt \sqrt{2}|d_0|$ which leads to
\begin{align}
\delE=
\frac{\mathcal{E}^2d^2}{8}
\left(
\frac{1+\tau\sin2\alpha}{\varepsilon_{cv}+\hbar\omega}+
\frac{1-\tau\sin2\alpha}{\varepsilon_{cv}-\hbar\omega}
\right)\,.
\end{align}
Since we are studying THG, we can use $\hbar\omega=\varepsilon_{cv}/3$ which simplifies the above equation into
\begin{align}
\delE=\frac{3\mathcal{E}^2d^2}{32\varepsilon_{cv}}(3-\tau\sin2\alpha)\,.\label{e49}
\end{align}

\section{S3 Quantum-mechanical model for Third Harmonic \mbox{Generation}}

\subsection{General analytical expression for the $\mathbf{\chi^{(3)}}$ tensor of a  solid}
In this section, we focus on analytical expressions for the elements of the third-order susceptibility, $\chi_{dcba}^{(3)}(-\omega_3;\omega_{\gamma},\omega_{\beta},\omega_{\alpha})$ with $d$, $c$, $b$, $a$ $\in\{x,y\}$ where $\omega_\alpha$, $\omega_\beta$, and $\omega_\gamma$ are the incoming frequencies and $\omega_3=\omega_\alpha+\omega_\beta+\omega_\gamma$ is the frequency of the outgoing four-wave mixing signal.
A general analytical expression for the third order optical susceptibility has been derived in Ref.~\cite[equation~(35)]{Aversa1995}:
\begin{align}\hspace{-0.7em}
\frac{\chi_{dcba}^{(3)}}{C}=&\hspace{-0.7em} \sum_{l, m, n, p, \mathbf{k}} \frac{r_{m n}^d}{\varepsilon_{n m}-\omega_3}\left[\frac{r_{n l}^c}{\varepsilon_{l m}-\omega_2}\left(\frac{r_{l p}^b r_{p m}^a f_{m p}}{\varepsilon_{p m}-\omega_1}-\frac{r_{l p}^a r_{p m}^b f_{p l}}{\varepsilon_{l p}-\omega_1}\right)-\left(\frac{r_{n l}^b r_{l p}^a f_{p l}}{\varepsilon_{l p}-\omega_1}-\frac{r_{n l}^a r_{l p}^b f_{l n}}{\varepsilon_{n l}-\omega_1}\right) \frac{r_{p m}^c}{\varepsilon_{n p}-\omega_2}\right]\nonumber\\
+\,i& \sum_{l, m, n, \mathbf{k}} \frac{r_{m n}^d}{\varepsilon_{n m}-\omega_3}\left[\frac{1}{\varepsilon_{n m}-\omega_2}\left(\frac{r_{n l}^b r_{l m}^a f_{m l}}{\varepsilon_{l m}-\omega_1}-\frac{r_{n l}^a r_{l m}^b f_{l n}}{\varepsilon_{n l}-\omega_1}\right)\right]_{; k^c} \nonumber\\
+\,i& \sum_{l, m, n, \mathbf{k}} \frac{r_{m n}^d}{\varepsilon_{n m}-\omega_3}\left[\frac{r_{n l}^c}{\varepsilon_{l m}-\omega_2}\left(\frac{r_{l m}^a f_{m l}}{\varepsilon_{l m}-\omega_1}\right)_{; k^b}-\left(\frac{r_{n l}^a f_{l n}}{\varepsilon_{n l}-\omega_1}\right)_{; k^b} \frac{r_{l m}^c}{\varepsilon_{n l}-\omega_2}\right] \nonumber\\
-&\sum_{m, n, \mathbf{k}} \frac{r_{m n}^d}{\varepsilon_{n m}-\omega_3}\left[\frac{1}{\varepsilon_{n m}-\omega_2}\left(\frac{r_{n m}^a f_{m n}}{\varepsilon_{n m}-\omega_1}\right)_{; k^b}\right]_{; k^c}\label{e1}\,.
\end{align}
Here, $\varepsilon_{nm}$ is the difference between the energies of the bands $n$ and $m$ at crystal momentum~$\bk$, $\omega_1=\omega_\alpha$, $\omega_2=\omega_\alpha+\omega_\beta$, $\mathbf{r}_{nm}=(1-\delta_{nm})\mathbf{d}_{nm}$, where $\mathbf{d}_{nm}$ is the dipole matrix element which can be computed as integral over the Brillouin zone volume~$\Omega$ using the lattice-periodic functions~$u_{n\bk}(\br)$ ,
\begin{align}
\mathbf{d}_{nm}=\frac{(2\pi)^3i}{\Omega}\int_\Omega d^3\mathbf{r}\,u^*_{n\mathbf{k}}(\mathbf{r})\nabla _\bk\, u_{m\mathbf{k}}(\mathbf{r})\,.
\end{align}
In equation~\eqref{e1}, the generalized derivative $
(S_{nm})_{;\mathbf{k}}\equiv\frac{\partial S_{nm}}{\partial \mathbf{k}}-iS_{nm}(\mathbf{d}_{nn}-\mathbf{d}_{mm})$ of a general quantity~$S$ is used. 
We have suppressed all $\bk$-dependencies in equation~\eqref{e1}.

\subsection{Model Hamiltonian for monolayer TMDs}
We employ a two-band model applicable near the $\pm$K points of the Brillouin zone to represent the TMD Hamiltonian~\cite{Taghizadeh2019prb,Herrmann2025},
\begin{align}
    \boldsymbol{h}(\bk) = 
\begin{pmatrix}
     \Delta+\delE  & \,\gamma^* f^*(\bk) \\[0.5em]
    \,\gamma f(\bk) & -\Delta-\delE 
\end{pmatrix}\,,
     \label{e27}
\end{align}
where $\Delta$ is the on-site energy, $\gamma$ is an effective hopping, and 
\begin{align}
    f(\bk) &= \frac{2}{\sqrt{3}}e^{-i\frac{\pi}{3}}\left(e^{ik_xa/\sqrt{3}} + 2  e^{-ik_xa/(2\sqrt{3})} \cos(k_ya/2)\right)
\end{align}
with lattice constant $a$. The term $\delE$ is the time-reversal symmetry breaking term due to the optical Stark and Bloch-Siegert effects 
 at the $\pm$K points, $\pm$K$=\frac{2\pi}{3}(\sqrt{3},\tau)$, where $\tau=\pm1$ (see S2). 
As our driving pulses excite the TH resonantly at the $\pm$K points, we employ the Taylor expansion of $f(\bk)$ near $\pm$K,
\begin{align}
f(\bk)=-(i\kappa_x+\kappa_y\tau)
+\zeta_2( \kappa_x+i\kappa_y \tau)^2
+\zeta_3(i\kappa_x+\kappa_y\tau)(\kappa_x^2+\kappa_y^2)
+\mathcal{O}(\kappa^4)
\end{align}
with $\kappa_{x(y)}=a(k_{x(y)}-\text{K}_{x(y)})$, $
\zeta_2=\frac{\sqrt{3}}{12}$, $
\zeta_3=\frac{1}{24}$.

\subsection{Evaluating $\chi^{(3)}$ for a monolayer TMD}
For THG, \textit{i.e.} , $\omega_\gamma=\omega_\beta=\omega_\alpha=:\omega$,  a two band model with occupied valence band ($f_v=1$) and empty conduction band ($f_c=0$), equation~\eqref{e1} simplifies to
\begin{multline}
\frac{\chi^{(3)}_{dcba}}{C}= \mathcal{P}_I \sum_{\mathbf{k}} \frac{-d_{v c}^d(\bk) d_{c v}^c(\bk)}{(\varepsilon_{cv}(\bk)-3\hbar\omega+i\hbar/T_2)\hbar\omega}\left(\frac{d_{c v}^a(\bk) d_{v c}^b(\bk)}{\varepsilon_{cv}(\bk)-\hbar\omega}+\frac{d_{v c}^a(\bk) d_{c v}^b(\bk)}{\varepsilon_{v c}(\bk)-\hbar\omega}\right)\\
-\frac{d_{v c}^d(\bk)}{\varepsilon_{c v}(\bk)-3\omega+i\hbar/T_2}\;\frac{\partial}{\partial k_c}\left[\frac{1}{\varepsilon_{c v}(\bk)-2\omega}\;\frac{\partial}{\partial k_b}\left(\frac{d_{c v}^a(\bk) f_{v c}}{\varepsilon_{c v}(\bk)-\omega}\right)\right]\label{eq:chi2band}\,.
\end{multline}
Here, we have included the dephasing rate $1/T_2$~\cite{korolev2024unveilingroleelectronphononscattering,Herrmann2025} as it is commonly used to model ultrafast decoherence processes which include electron-phonon and electron-electron scattering. The symbol $\mathcal{P}_I$ denotes the intrinsic permutation operator, since $c$, $b$, $a$ are dummy indexes which can be freely exchanged in the case of THG. We have excluded terms with $\mathbf{d}_{vv}$ and $\mathbf{d}_{cc}$ from equation~\eqref{eq:chi2band} since they are quadratic in $\boldsymbol{k}$ for the model Hamiltonian (equation~\eqref{e27})~\cite{Herrmann2025}.

For TH resonant driving, $\varepsilon_{cv}\approx3\hbar\omega$ in equation~\eqref{eq:chi2band} at the $\pm $K points, $\varepsilon_{cv}(\pm \text{K})=2(\Delta +\delE)$, the effect of TRS breaking \textit{via} $\delE$  is dominant in the term 
\begin{align}
\frac{1}{\varepsilon_{cv}(\pm \text{K})+i\hbar/T_2-3\hbar\omega} = \frac{1}{2(\Delta +\delE)+i\hbar/T_2-3\hbar\omega} \,.   \label{e7a}
\end{align}

For evaluating equation~\eqref{eq:chi2band} besides the resonant term~\eqref{e7a}, we calculate the dipoles and the eigenstates of the unperturbed Hamiltonian ($\delE=0$). 
The eigenvalues close to $\pm $K are
\begin{align}
\varepsilon_v=-\sqrt{\Delta^2+\abs{\gamma f}^2}\;\,\overset{\text{at $\pm$K}}{=}-\Delta,\quad \varepsilon_c=\sqrt{\Delta^2+\abs{\gamma f}^2}\;\,\overset{\text{at $\pm$K}}{=}\Delta,\quad 
\varepsilon_{cv}\overset{\text{at $\pm$K}}{=} 2\Delta\,. \label{eq:eigenvalues}
\end{align}
and the eigenstates are
\begin{align}
\ket{v\mathbf{k}}=\frac{1}{N(\mathbf{k})}
\begin{pmatrix}
-\gamma^*f^*(\mathbf{k} )\\
\Delta+\sqrt{\Delta^2+\abs{\gamma f(\mathbf{k})}^2}
\end{pmatrix},\quad
\ket{c\mathbf{k}}=\frac{1}{N(\mathbf{k})}
\begin{pmatrix}
\Delta+\sqrt{\Delta^2+\abs{\gamma f(\mathbf{k})}^2}\\
\gamma f(\mathbf{k} )
\end{pmatrix}\,, \label{e7}
\end{align}
where
\begin{align}
N(\mathbf{k})=\sqrt{\left(\Delta+\sqrt{\Delta^2+\abs{\gamma f(\mathbf{k})}^2}\right)^2+\abs{\gamma f}^2}\;\,\overset{\text{at $\pm$K}}{=}2\Delta.
\end{align}
Therefore,
\begin{align}
\ket{v\mathbf{k}}=
\begin{pmatrix}
\displaystyle-\frac{\gamma^*f^*(\mathbf{k} )}{2\Delta}\\[0.5em]
1
\end{pmatrix},\quad
\ket{c\mathbf{k}}=
\begin{pmatrix}
1\\[0.5em]
\displaystyle \frac{\gamma f(\mathbf{k} )}{2\Delta}
\end{pmatrix}\,.
\end{align}
Thus, the dipole matrix elements are
\begin{align}
d^x_{cv}&=iq\bra{c\bk}\frac{\partial}{\partial k_x}\ket{v\bk}=\frac{a q \gamma ^*}{2 \Delta }(1-2 i  \zeta _2 \kappa_x-2 \zeta _2 \tau  \kappa_y-3 \zeta _3 \kappa_x^2-2 i \zeta _3 \tau  \kappa_x \kappa_y-\zeta _3 \kappa_y^2)+\mathcal{O}(\kappa^3)\,,\nonumber\\
d^y_{cv}&=iq\bra{c\bk}\frac{\partial}{\partial k_y}\ket{v\bk}=\frac{a q \gamma ^*}{2 \Delta }(i \tau-2\zeta _2 \tau  \kappa_x+2 i \zeta _2 \kappa_y-i  \zeta _3 \tau  \kappa_x^2-2  \zeta _3 \kappa_x \kappa_y-3 i  \zeta _3 \tau  \kappa_y^2)+\mathcal{O}(\kappa^3)\,,\nonumber\\
d^x_{vc}&=iq\bra{v\bk}\frac{\partial}{\partial k_x}\ket{c\bk}=\frac{a q \gamma }{2 \Delta }(1+2 i  \zeta _2 \kappa_x-2  \zeta _2 \tau  \kappa_y-3  \zeta _3 \kappa_x^2+2 i  \zeta _3 \tau  \kappa_x \kappa_y- \zeta _3 \kappa_y^2)+\mathcal{O}(\kappa^3)\,,\nonumber\\
d^y_{vc}&=iq\bra{v\bk}\frac{\partial}{\partial k_y}\ket{c\bk}=\frac{a q \gamma }{2 \Delta }(-i \tau-2 \zeta _2 \tau  \kappa_x-2 i \zeta _2 \kappa_y+i  \zeta _3 \tau  \kappa_x^2-2  \zeta _3 \kappa_x \kappa_y+3 i  \zeta _3 \tau  \kappa_y^2)+\mathcal{O}(\kappa^3)\,,\label{e11}
\end{align}
where $q$ is the charge of an electron. We evaluate $\chi^{(3)}_{dcba}$ from equation~\eqref{eq:chi2band} at the resonant $\pm $K points using the band energies~\eqref{eq:eigenvalues},  the dipoles~\eqref{e11} and the third-harmonic resonance~\eqref{e7a},

\begin{align}
\chi^{(3)}_{xxxx}       =\sum_{\tau=\pm1}&\chi^{(3)}_{xxxx}(\tau)
=\sum_{\tau\pm1}\frac{\mathcal{C}}{2(\Delta+\delE)-3\hbar\omega+{i\hbar}/{T_2}}
=\frac{2\detuning}{\detuning^2-(\deltagap)^2}\,\mathcal{C}=\chiint\label{e34}
\end{align}
with
\begin{align}
\detuning&:=2\Delta-3\hbar\omega+\frac{i\hbar}{T_2}+\sum_{\tau=\pm 1} \delE\overset{\eqref{e49}}{=}2\Delta-3\hbar\omega+\frac{i\hbar}{T_2}+\frac{9\mathcal{E}^2d^2}{32\Delta}\,, \\[0.5em]
\deltagap&:=\sum_{\tau=\pm 1} (-\tau) \cdot \delE
=\mathbf{\Delta} E_{ -K,\alpha}-\mathbf{\Delta} E_{ +K,\alpha}
\overset{\eqref{e49}}{=}\frac{3\mathcal{E}^2d^2}{32\Delta}\sin2\alpha\,,
\\[0.5em]
\mathcal{C}&:=
C\frac{a^4 q^4 \abs{\gamma}^2 \left(\abs{\gamma}^2 \left(2 \Delta ^2+5 \Delta  \hbar\omega -\hbar\omega ^2\right)+6 \Delta ^2 \zeta _3 \left(8 \Delta ^2-\hbar\omega ^2\right)\right)}{4 \Delta ^4(\Delta -\hbar\omega ) (2 \Delta-\hbar\omega)^2 (2 \Delta +\hbar\omega )}\,.
\end{align}
Similarly, the expression for $\chi^{(3)}_{xyyy}$ reads
\begin{align}
\chi^{(3)}_{xyyy}       =\sum_{\tau=\pm1}&\chi^{(3)}_{xyyy}(\tau)
=\sum_{\tau=\pm1}\frac{i\tau\mathcal{C}}{2(\Delta+\delE)-3\hbar\omega+{i\hbar}/{T_2}}
=\frac{2i\deltagap}{\detuning^2-(\deltagap)^2}\,\mathcal{C}=\chitrs \,.\label{e38}
\end{align}
Moreover, it is possible by evaluating equation~\eqref{eq:chi2band} to validate that the relations for the  $\bar6m'2'$ magnetic point group are fulfilled (detailed calculation not shown):
\begin{align*}
    \chi^{(3)}_{xxxx} &= \chi^{(3)}_{yyyy} = 3\chi^{(3)}_{xxyy} =3 \chi^{(3)}_{xyyx} =3 \chi^{(3)}_{xyxy} \equiv \chiint\\[0.5em]
    \chi^{(3)}_{xyyy} &=3\chi^{(3)}_{xxxy} = 3\chi^{(3)}_{xxyx} = 3\chi^{(3)}_{xyxx} = -3\chi^{(3)}_{yyyx} = -3\chi^{(3)}_{yyxy} = -3\chi^{(3)}_{yxyy}  = -\chi^{(3)}_{yxxx}\equiv \chitrs\,.
\end{align*}

\subsection{Evaluation of the third-harmonic rotation from the $\mathbf{\chi^{(3)}}$ tensor}
For evaluating the polarization rotation of the TH, we start from the third-order polarization,
\begin{align}
&P_d=\varepsilon_0\sum_{a,b,c}\chi_{dcba}^{(3)}E_cE_bE_a
\\&=\varepsilon_0\left[\chi^{(3)}_{dxxx}E_x^3+(\chi^{(3)}_{dxxy}+\chi^{(3)}_{dxyx}+\chi^{(3)}_{dyxx})E_x^2E_y+(\chi^{(3)}_{dxyy}+\chi^{(3)}_{dyyx}+\chi^{(3)}_{dyxy})E_xE_y^2+\chi^{(3)}_{dyyy}E_y^3\right]\,.\label{eqn:pxexp}
\end{align}
Focusing on the x-component of the emission ($d=x$), we know from the symmetry of the $\bar6m'2'$ magnetic point group that
\begin{align*}
\chi^{(3)}_{xxxy}+\chi^{(3)}_{xxyx}+\chi^{(3)}_{xyxx}=\chi^{(3)}_{xyyy}\hspace{2em}\text{and}\hspace{2em}
\chi^{(3)}_{xxyy}+\chi^{(3)}_{xyyx}+\chi^{(3)}_{xyxy}=\chi^{(3)}_{xxxx}\,.
\end{align*}
As driving field, we employ equation~\eqref{e7f} without the phase factor $ \exp{ i\frac{\pi}{4}} $,
\begin{align}
    \boldsymbol{E}_\mathrm{IN} = \left(
   \begin{array}{c}
E_x \\ E_y
   \end{array}
   \right)
    = \mathcal{E} 
    \left(
   \begin{array}{c}
\cos\alpha \\  i \sin \alpha
   \end{array}
   \right)
   .\label{e7fa}
\end{align}
With this, equation (\ref{eqn:pxexp}) for emission along the x direction ($d=x$) becomes
\begin{align}
P_x&=\varepsilon_0\left(\chi^{(3)}_{xxxx}E_x^3+\chi^{(3)}_{xyyy}E_x^2E_y+\chi^{(3)}_{xxxx}E_xE_y^2+\chi^{(3)}_{xyyy}E_y^3\right)\nonumber\\
&=\varepsilon_0\chi^{(3)}_{xxxx}E_x(E_x^2+E_y^2)+\varepsilon_0\chi^{(3)}_{xyyy}E_y(E_x^2+E_y^2)\nonumber\\
&=\varepsilon_0(E_x^2+E_y^2)(\chiint E_x+\chitrs E_y)\nonumber\\
&\overset{\eqref{e7fa}}{=}\varepsilon_0\mathcal{E}^3(\cos^2\alpha-\sin^2\alpha)(\chiint\cos\alpha+i\chitrs\sin\alpha)\nonumber\\
&=\varepsilon_0\mathcal{E}^3\cos2\alpha(\chiint\cos\alpha+i\chitrs\sin\alpha)\label{eqn:px}\,.
\end{align}
Similarly,
\begin{align}
P_y&=\varepsilon_0\left(\chi^{(3)}_{yyyy}E_y^3+\chi^{(3)}_{yxxx}E_y^2E_x+\chi^{(3)}_{yyyy}E_yE_x^2+\chi^{(3)}_{yxxx}E_x^3\right)\nonumber\\
&=\varepsilon_0\chi^{(3)}_{yyyy}E_y(E_x^2+E_y^2)+\varepsilon_0\chi^{(3)}_{yxxx}E_x(E_x^2+E_y^2)\nonumber\\
&=\varepsilon_0(E_x^2+E_y^2)(\chi^{(3)}_{yxxx}E_x+\chi^{(3)}_{yyyy}E_y)\nonumber\\
&\overset{\eqref{e7fa}}{=}\varepsilon_0\mathcal{E}^3(\cos^2\alpha-\sin^2\alpha)(-\chitrs\cos\alpha+i\chiint\sin\alpha)\nonumber\\
&=-\varepsilon_0\mathcal{E}^3\cos2\alpha(\chitrs\cos\alpha-i\chiint\sin\alpha)\label{eqn:py}\,.
\end{align}
Using equations (\ref{eqn:px}) and (\ref{eqn:py}), the Stokes parameters can be calculated as,
\begin{align}
S_1&\propto\abs{P_x}^2-\abs{P_y}^2\nonumber\\
&=\mathcal{E}^6 \cos ^2 2\alpha [(\chiint \cos \alpha+i \chitrs \sin \alpha)(\chiint^* \cos \alpha-i   \chitrs^* \sin \alpha)\nonumber\\
&-(\chitrs \cos \alpha-i  \chiint \sin \alpha) (\chitrs^* \cos \alpha+i   \chi_1^* \sin \alpha)]\nonumber\\
&=\mathcal{E}^6 \cos ^3 2\alpha \left(\abs{\chiint}^2-\abs{\chitrs}^2\right)\\
S_2&\propto\Re{P_xP_y^*}\nonumber\\
&=-\mathcal{E}^6 \cos ^2 2\alpha[(\chiint \cos \alpha+i   \chitrs \sin \alpha)(\chitrs^* \cos \alpha+i   \chiint^* \sin \alpha)\nonumber\\
&+(\chiint^* \cos \alpha-i   \chitrs^* \sin \alpha)(\chitrs \cos \alpha-i  \chiint \sin \alpha) ]\nonumber\\
&=-\mathcal{E}^6 \cos ^3 2\alpha \left(\chiint\chitrs^*+\chitrs\chiint^*\right)\,.
\end{align}

From this, the angle of rotation of the ellipse $\theta$ can be calculated as
\begin{align}
\tan2\theta&=\frac{S_2}{S_1}
=-2\frac{\text{Re}(\chiint\chitrs^*)}{\abs{\chiint}^2-\abs{\chitrs}^2}
\overset{\eqref{e34},\eqref{e38}}{=} \frac{-2\deltagap\,\text{Im}\detuning }{|\detuning|^2-|\deltagap|^2}\label{eq:THG final}
\\[0.5em]&
= \frac{-\frac{3\mathcal{E}^2d^2}{16\Delta}\sin2\alpha\,(\hbar/T_2)}{\left(2\Delta-3\hbar\omega+\frac{9\mathcal{E}^2d^2}{32\Delta}\right)^2+\left(\hbar/T_2\right)^2-\left(\frac{3\mathcal{E}^2d^2}{32\Delta}\right)^2\sin^2 2\alpha}\,.
\end{align}
For a small electric field, $\mathcal{E}^2d^2/(\hbar\Delta/T_2)\ll 1$, the TRS breaking~$|\deltagap|\ll |\detuning|$ and, thus, the rotation of the TH signal is linear in $\deltagap$ and the intensity:
\begin{align}
\tan 2\theta&=
-\, \frac{2}{1+(2\Delta-3\hbar\omega)^2/(\hbar/T_2)^2}\;\frac{\deltagap}{\hbar/T_2}
\\[0.5em]
&=
-\, \frac{3}{16}\,\sin2\alpha\;\frac{1}{1+(2\Delta-3\hbar\omega)^2/(\hbar/T_2)^2}\;\frac{\mathcal{E}^2d^2}{\hbar\Delta/T_2}\,. \label{e8}
\end{align}

\section{S4 Stereographic projection}
\captionsetup{format=plain,justification=raggedright}

\begin{figure}
\centering
    \includegraphics[width=\linewidth]{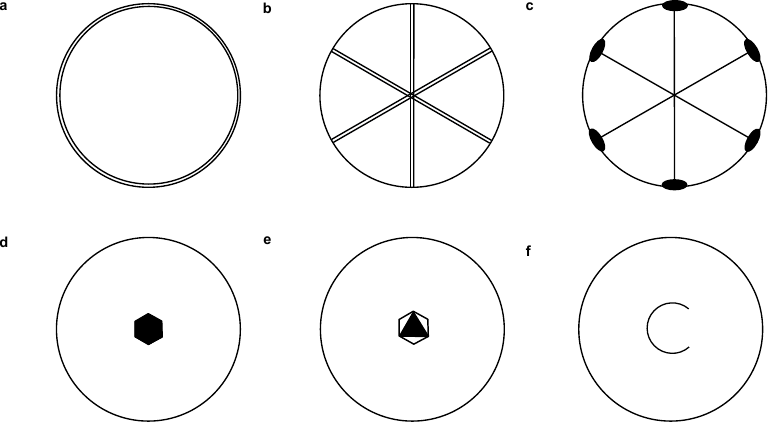}
    \caption{\textbf{Stereographic projection of different point symmetry operations} Each element \textbf{a} to \textbf{f} represents the stereographic projection of a specific symmetry operation. \newline \textbf{a} Horizontal mirror. \textbf{b} Vertical mirror. \textbf{c} $C_2$ operation, \textit{i.e.} rotation along a vertical axis. \textbf{d} n-fold rotation symmetry (here $n=6$). \textbf{e} The system has x-fold symmetry, but only n-fold rotation symmetry (here $x=6$ and $n=3$). \textbf{f} Roto-inversion (rotation combined with inversion).}
    \label{fig:SI_FIG_Stereographic_Projection}
\end{figure}

In this section, we discuss the stereographic projection that we use in the main text to identify symmetry operations for crystallographic point groups and magnetic groups in TMDs. Relevant symmetry operations and their stereographic projections are shown in Fig.~\ref{fig:SI_FIG_Stereographic_Projection}, while a more detailed explanation can be found \textit{e.g.} in Ref.~\cite{Dresselhaus2007}. We show six exemplary symmetry operations: horizontal mirror (Fig.~\ref{fig:SI_FIG_Stereographic_Projection}a), vertical mirror (Fig.~\ref{fig:SI_FIG_Stereographic_Projection}b), $C_2$ rotation (rotation along vertical axis vector, Fig.~\ref{fig:SI_FIG_Stereographic_Projection}c), $n$-fold rotational symmetry (here 6-fold, Fig.~\ref{fig:SI_FIG_Stereographic_Projection}d), existing x-fold symmetry in the system, but only n-fold rotation symmetry (Fig.~\ref{fig:SI_FIG_Stereographic_Projection}e) and roto-inversion (Fig.~\ref{fig:SI_FIG_Stereographic_Projection}f). One can combine the given symmetry operations into one single stereographic projection to obtain the graphs shown in Fig.~2b, e, h and k of the main text. For example, a TMD monolayer with preserved TRS belongs to the $\bar6m2$ symmetry group, and thus the allowed symmetry operations are (Fig.~2b of the main text): one horizontal mirror, three vertical mirrors, 3-fold rotation symmetry and three $C_2$ rotations. When TRS is broken, the symmetry of monolayer TMDs is reduced to the magnetic group $\bar6m'2'$, and some symmetry operations are allowed only in combination with the antisymmetry operation (Fig.~2e of the main text): $C_2$ rotations and vertical mirror planes are thus depicted in red. A similar situation appears in TMD bilayers, that belong to the $\bar3m$ group (Fig.~2h of the main text) when TRS is preserved. Here, the symmetry operations are: three $C_2$ rotations, three vertical mirrors, 3-fold rotation symmetry and roto-inversion. When TRS is broken, the symmetry of TMD bilayers is reduced to the $\bar3m'$ magnetic group (Fig.~2k of the main text), where only the 3-fold rotation symmetry and roto-inversion are allowed without the antisymmetry operation.